\documentclass[12pt,preprint]{aastex}

\begin{document}

\title{Unveiling supergiant fast X-ray transient sources with INTEGRAL}

\author{ Sguera V.\altaffilmark{a}, Bazzano A.\altaffilmark{b}, 
   Bird A. J.\altaffilmark{a},
   Dean A. J.\altaffilmark{a},    
   Ubertini P.\altaffilmark{b},
   Barlow  E. J. \altaffilmark{a}, 
   Bassani L.\altaffilmark{c},
   Clark D. J. \altaffilmark{a},
   Hill A. B. \altaffilmark{a},
   Malizia A. \altaffilmark{c},
   Molina  M. \altaffilmark{a},
   Stephen J. B. \altaffilmark{c}
   }

\altaffiltext{a}{School of Physics and Astronomy, University of
Southampton, SO17 1BJ, UK}
\altaffiltext{b}{IASF/INAF, Rome, Italy}
\altaffiltext{c}{IASF/INAF, Bologna, Italy}

\begin{abstract}
Supergiant high mass X-ray binaries (SGXBs) are believed to be rare objects,  
as stars in the supergiant phase have a very short lifetime and  to date only about a dozen of them have been discovered. They are known to be 
persistent and bright X-ray sources.    
INTEGRAL is  changing  this  classical picture as its observations are revealing the presence of a new 
subclass of SGXBs which have been labelled as supergiant fast X-ray transients (SFXTs) as they are strongly characterized 
by fast X-ray outbursts lasting less than a day, typically a few hours. We report on  IBIS detections of  
newly discovered fast X-ray outbursts from 10 sources, four of which  
have been recently optically identified as supergiant high mass X-ray binaries. In particular for one of them, IGR~J11215$-$5952,
we observe   fast X-ray transient behaviour for the first time.
The remaining six sources 
(IGR~J16479$-$4514, IGR~J16418$-$4532, IGR~J16195$-$4945=AX~J161929$-$4945, XTE~J1743$-$363,
AX~J1749.1$-$2733, IGR~J17407$-$2808) 
are still unclassified, however they can be considered as candidate SFXTs
because of their similarity to the known SFXTs. 
\end{abstract}

\date{Received 20 December 2005 / Accepted 27 March 2006}


\section{Introduction} 
High mass X-ray binaries (HMXBs) are systems composed of an accreting compact object (magnetized neutron star or black hole) and 
an early-type massive star (White et al. 1995).  
The majority of the known HMXBs ($\sim$80\%) are Be/X-ray binaries consisting of a neutron star orbiting around a Be star;
the compact object is characterized by an eccentric orbit,  spending  most of its time far away from the disk surrounding the Be star. 
During the time of the neutron star's periastron passage, the compact object accretes from the dense equatorial Be star disk producing  
bright outbursts lasting for several weeks or even months. The other major group of HMXBs consists of a compact object orbiting around
a supergiant early-type star (OB) and in this case the X-ray emission is powered by accretion of material originating from the donor star through a strong 
stellar wind and/or  Roche-lobe overflow. 

Stars in the supergiant phase have a very short lifetime; due to the evolutionary timescales involved 
supergiant HMXBs (SGXBs) are expected to be much 
less numerous than Be/X ray binaries. To date about a dozen of them have been detected in the classical X-ray band.

INTEGRAL is changing  this  picture since its observations are indicating the presence of a new 
subclass of SGXBs which have been labelled as supergiant fast X-ray transients (SFXTs) (Negueruela et al. 2005a). 
Most of the time they are undetectable, then occasionally they undergo  fast X-ray transient activity 
lasting less than a day, typically a few hours (Sguera et al. 2005). Their outbursts show complex structures characterized 
by several fast flares with both rise and decay times of less than
1 hour (typically a few tens of minutes).  
This kind of X-ray transient activity is very different from that seen in other high mass X-ray binaries (i.e. Be/X-ray transients).
Moreover, SFXTs differ from classical SGXBs 
since the latter are known to be persistent bright sources with  X-ray luminosities always detectable in the range 10$^{35}$--10$^{36}$ erg s$^{-1}$
while, on the contrary, SFXTs present quiescence luminosities with values 
or upper limits in the range 10$^{32}$--10$^{33}$  erg s$^{-1}$ (Negueruela et al. 2005a).
The physical origin of the fast outbursts displayed by SFXTs is still unknown. Their very short duration  is not compatible with 
viscous timescales in a typical accretion disc; therefore fast X-ray outbursts from SFXTs must be due to a completely different mechanism.
in't Zand (2005) has suggested that the origin must be related to the early-type
donor star and in particular to the wind accretion mass transfer mode from the supergiant to the compact object. It could be that the supergiant ejects material
in a non-continuous way. In fact, many early-type stars are characterized by
highly structured and variable massive winds (Prinja et al. 2005)
which could have a fundamentally clumpy nature. The capture of these clumps
by  a nearby compact object could then produce fast X-ray flares thereby explaining the observed timescales.

SFXTs are difficult to detect because of their very transitory nature; 
to date the list consists of 5 objects. However there are a few more unidentified X-ray sources which 
display  fast X-ray outbursts and are therefore  candidate  SFXTs even if their optical counterpart has not yet been 
identified with an early-type supergiant (Negueruela et al. 2005a). 

The IBIS/ISGRI instrument  (Ubertini et al. 2003, Lebrun et al. 2003) on board the INTEGRAL satellite (Winkler et al. 2003) is 
particularly suited to the detection of new or already known supergiant  fast X-ray transient sources 
since it provides a very powerful combination of  a large FOV,  good sensitivity
and wide energy range. All are  indispensable characteristics for the detection
of fast transient events such as SFXTs.

In this paper, we report in section 3 on IBIS detections of newly discovered fast X-ray 
outbursts from 3 known SFXTs (XTE~J1739$-$302,  IGR~J18410$-$0535=AX~J1841.0$-$0536,
IGR~J17544$-$2619). Moreover, we unveil for the first time the fast X-ray transient nature of IGR~J11215$-$5952 which has been recently 
identified as a supergiant HMXB (Masetti et al. 2005). 
Furthermore, we report in section 4 on new IBIS discoveries  of fast X-ray outbursts 
from 6 more sources (IGR~J16479$-$4514, IGR~J16418$-$4532, AX~J1749.1$-$2733,
IGR~J16195$-$4945=AX~J161929$-$4945, XTE~J1743$-$363,
IGR~J17407$-$2808). All of them are still unidentified, but they can be considered as  candidate SFXTs
since their characteristics strongly resemble those of the  known SFXTs.  
\section{INTEGRAL data analysis}
The data are collected with the low-energy array, ISGRI (INTEGRAL Soft Gamma-Ray Imager; Lebrun et al. 2003), consisting of a pixilated 128$\times$128 
CdTe solid-state detector that views the sky through a coded aperture mask.
The reduction and analysis of the ISGRI data have been performed by using the INTEGRAL Offline Scientific Analysis (OSA) v.4.2 available
to the public through the INTEGRAL Science Data Centre ISDC (Courvoisier et al. 2003). 
INTEGRAL observations are typically divided into short pointings (Science Window, ScW) of $\sim$ 2000 s duration.
Our ScW dataset belongs to the Core Program data collected as part of the Galactic Plane Survey (GPS) and
the Galactic Centre Deep Exposure (GCDE) (Winkler et al. 2003) from revolution 45 (end of February 2003)
to 307  (April 2005) as well as to all public data released up to revolution 160.

The set of observations used for our analysis, although being highly inhomogeneous in both extent and exposure,
covers  $\sim$65\% of the sky at a level of at least 10 ksec,
as can be seen from the exposure map shown in Figure 1. The Galactic Plane, and in particular the Galactic Center,
are particularly well covered with exposure times of at least 100 and 1000 ksec respectively.
It is not surprising that most of the SFXTs so far discovered by INTEGRAL are located in the proximity 
of the Galactic Centre, a region that has been extensively
monitored.

\clearpage
\begin{figure}[t!]
\plotone{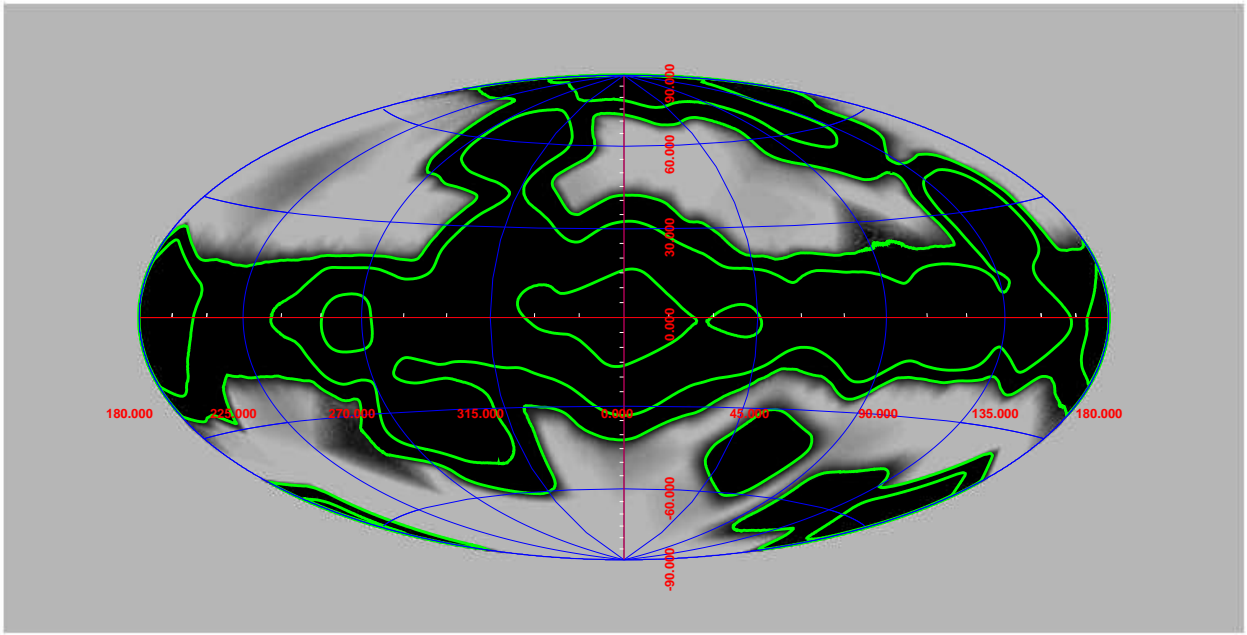}
\caption{The exposure map for our dataset. Contour levels are at 10, 100, 1000 ksec.}
\end{figure}
\clearpage
 
Figure 2 displays the angular distribution off the Galactic plane 
of the central pointing direction of all ($\sim$ 11300) ScWs in our dataset. Their strong concentration 
($\sim$70\%, 7900 ScWs) towards the Galactic plane ($|$b$|$$\leq$ 5$^{\circ}$) rather than medium or high galactic latitude is clearly evident.
SFXTs are expected to be mainly located on the Galactic plane due to the very young ages of their progenitor stars.
\begin{figure}[t!]
\plotone{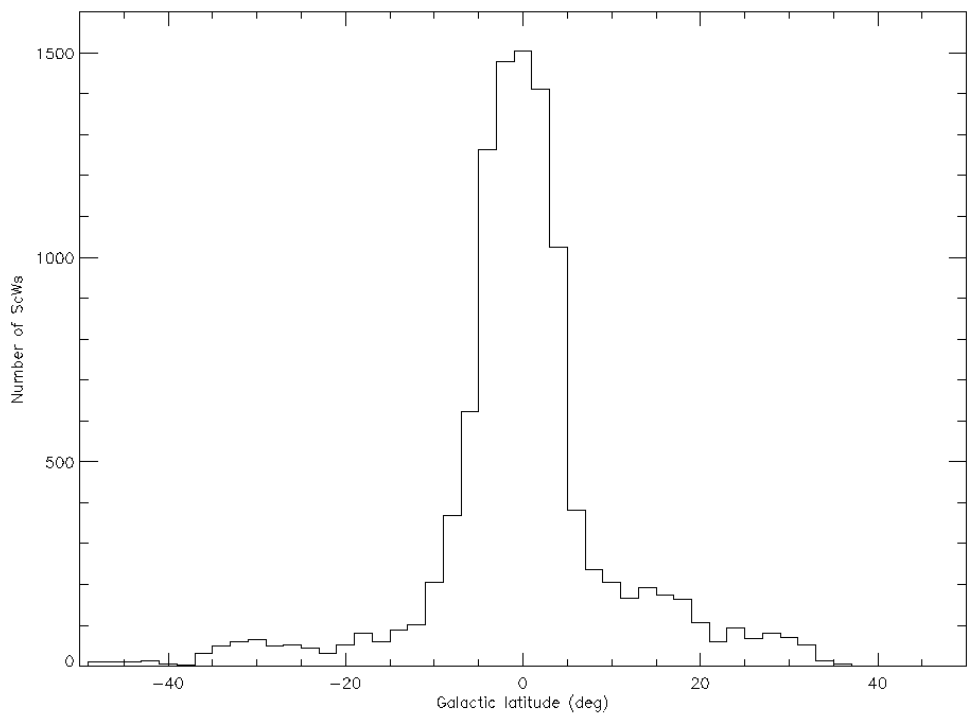}
\caption{The number of ScWs versus Galactic latitude of their pointing.}
\end{figure}

\clearpage
Each ScW of our dataset has been analysed using the imaging components of the OSA4.2. A proprietary tool
was used to search each ScW, and provide a cross-correlated list of all excesses above 5$\sigma$ found in one or more ScWs.
The search was initially performed in the energy band  20--30 keV; but
when a  source was found then we also checked the detection at higher energies and in the previous and subsequent ScWs. 
This approach is efficient  in unveiling very fast transient events lasting only a few hours since the search occurs on the same 
timescale as the outbursts themselves. Integrating for longer just degrades the signal-to-noise of the detection.
The resulting excesses were then inspected visually to ensure an appropriate point spread function and so to identify them as real sources. 
This was necessary in order to reject false detections such as ghosts, structures or background noise. 
For each newly discovered fast outburst a more detailed timing and spectral analysis was performed  using the OSA v.4.2. 

The sensitivity limit for a persistent source detected at  5$\sigma$ level (20--30 keV)  in only one ScW of $\sim$ 2000 s
is $\sim$ 25 mCrab. In the case of an outburst visible in more than one consecutive ScW, we can assume our detection limit 
is again  $\sim$ 25 mCrab, but now refers  to the average flux of the outburst during the ScW containing the peak. 
Note that we do not place any requirements on detection during multiple ScWs.

Due to possible cross-talk between objects in the same FOV,   
we have also investigated the variability pattern of all other bright sources in the FOV, 
as well as  the source of  interest. They have shown a different time variability enabling us to conclude that the light curves  obtained 
for  the  sources of  interest are reliable.
Images from the X-ray Monitor JEM-X (Lund et al. 2003) were created for all newly discovered outbursts reported in this paper. 
In most cases, the source was outside or on the edge of the JEM-X field of view; 
no significant detection were obtained for those inside the FOV so we did not take into account JEM-X data for our analysis.

\section{Firm Supergiant Fast X-ray Transients}

\subsection{XTE~J1739$-$302=IGR~J17391$-$3021}
\subsubsection{Archival X-ray observations of the source}
XTE~J1739$-$302=IGR~J17391$-$3021  is a well studied  SFXT (Smith et al. 2005). Since its discovery in August 1997 by RXTE
(Smith et al. 1998), the source has been detected in outburst by ASCA in 1999 (Sakano et al. 2002) and by INTEGRAL in 2003 and 2004 (Sguera et al. 2005, Lutovinov
et al. 2005). 
All previous reported outbursts are characterized by durations ranging from $\sim$ 2 hours to $\sim$ half a day. 
Sguera et al. (2005) showed several detailed ISGRI lightcurves of the source in outburst which clearly display complex structures with  quick flares 
reaching their peak flux over very short timescales (a few tens of minutes) and then dropping off with the same timescale as the rise time. 
To date, no pulsations have been reported from any of the outbursts detected by the previously cited X-ray missions.

Negueruela et al. (2005b) identified the optical counterpart as an O8Iab(f) supergiant at a distance of $\sim$ 2.3 kpc. In the optical and infrared, 
XTE~J1739$-$302 is very similar to the supergiant HMXB systems.
\subsubsection{Analysis of IBIS/ISGRI observations and results}
Table 1 contains a list of all IBIS detections of outbursts from XTE~J1739$-$302. 
Here we report on a newly discovered fast X-ray outburst of XTE~J1739$-$302 detected on 21 August 2004 (No. 6 in Table 1). 
Here, and in the subsequent analysis, we assume the beginning of the first ScW during which the source was detected as being  the start time of the outburst 
and similarly the burst stop time to be the end of the last ScW during which the source was detected.

The source turned on at  05:25:12 UTC  an showed abrupt outburst activity lasting $\sim$ 3 hours, then it
turned off at  08:38:03 UTC. As we can see from the 20--60 keV ISGRI
light curve shown in Figure 3, the fast X-ray transient activity is quite complex.  It is characterized by several sharp flares which reach their peak 
on very short timescale and then drop with a similar timescale. This kind of transient behaviour 
has already been noted in several other outbursts of XTE~J1739$-$302 detected by IBIS (Sguera et al. 2005).
A Fourier analysis of  the lightcurve in Figure 3 did not show any significant evidence of pulsations; however
we note that the peaks are approximately separated by $\sim$ 2000 s; another  
outburst detected by IBIS on March 2003 also displayed this kind of characteristic (Sguera et al. 2005). Unfortunately in both cases the duration of 
the outburst is only a few times the putative period of 2000~s,  so the detection of a longer outburst is needed to confirm
any such periodicity in the light curve of XTE~J1739$-$302.
To date this is the strongest outburst detected by IBIS from XTE~J1739$-$302.
The peak-flux and luminosity (20--60 keV) of the strongest flare in Figure 3 are $\sim$ 480 mCrab and  $\sim$3.5$\times$10$^{36}$ erg s$^{-1}$, respectively.  
The lowest limit for the quiescence luminosity (2--10 keV) of XTE~J1739$-$302 is  7$\times$10$^{32}$ erg s$^{-1}$ (Sakano et al. 2002, Smith et al. 2005), 
obtained from the same observation during which the source underwent a strong outburst.

\clearpage
\begin{figure}[t!]
\plotone{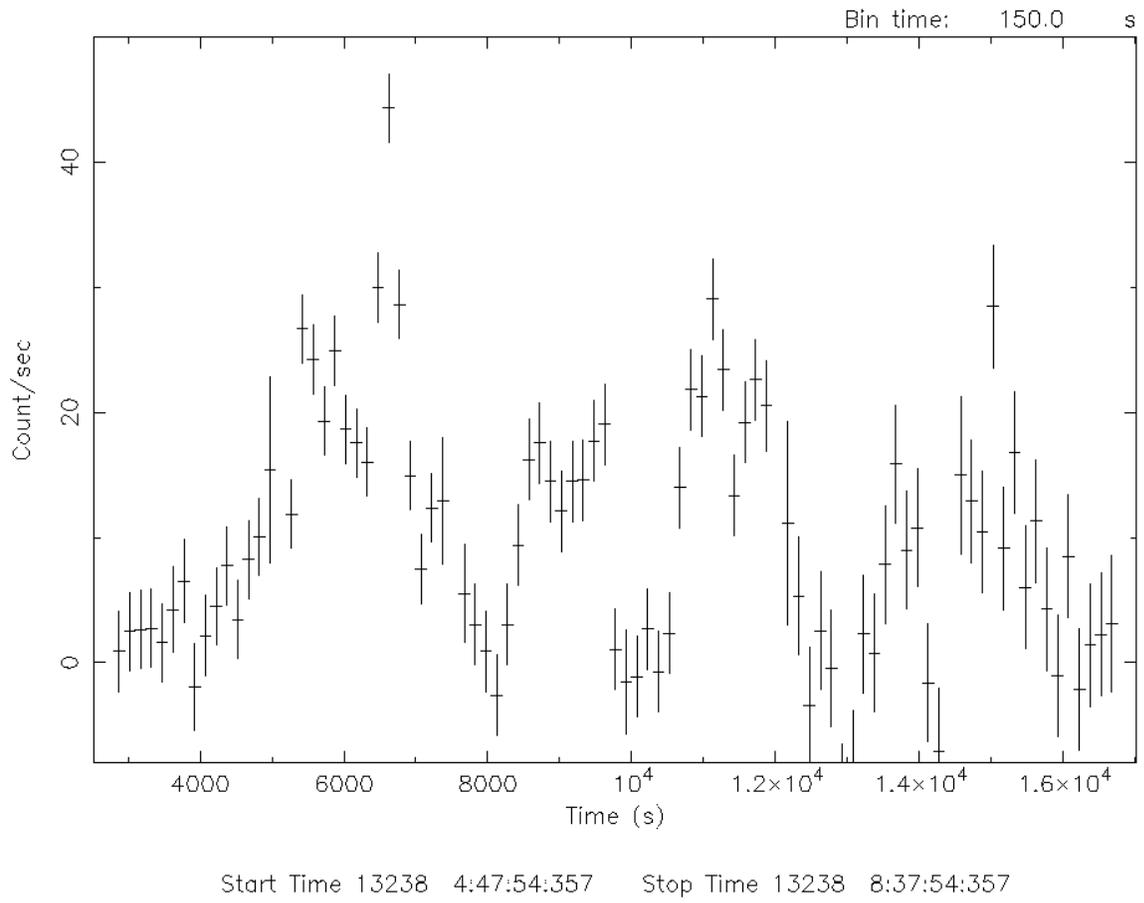}
\caption{The ISGRI light curve (20--60 keV) of a newly discovered outburst of XTE~J1739$-$302=IGR~J17391$-$3021}
\end{figure}
\clearpage                   
\begin{table*}[t!]
\begin{center}
\caption {Summary of ISGRI observations of outbursts of XTE~J1739$-$302=IGR~J17391$-$3021.}
\begin{tabular}{clcccc}
\hline
\hline
No.  & Date  & duration (hours) & flux at the peak & ref\\
\hline
1  & 22 March  2003 & $\sim$ 2  & $\sim$ 250 mCrab (20--30 keV)  & [1] \\
2  & 26 August 2003 & $\sim$ 14 & $\sim$ 120 mCrab (18--60 keV) & [2] \\  
3  & 6 September 2003 &  $\sim$ 7 &  $\sim$ 60 mCrab (18--60 keV)& [2] \\
4  & 9 March 2004 &  $\sim$ 0.5 &  $\sim$ 150 mCrab (20--30 keV)  & [1]  \\
5  & 10 March 2004 &  $\sim$ 1.5 &  $\sim$ 250 mCrab (20--30 keV)  & [1] \\
6  & 21 August 2004  &  $\sim$ 3 &  $\sim$ 480 mCrab (20--60 keV)  & $\star$  \\
\hline
\hline
\end{tabular}
\end{center}
[1] Sguera et al. 2005 [2] Lutovinov et al. 2005 \\ $\star$ = this paper \\
\end{table*} 
\clearpage

\subsection{IGR~J17544$-$2619}

\subsubsection{Archival X-ray observations of the source}
IGR~J17544$-$2619 is a fast X-ray transient source discovered with INTEGRAL on 17 September 2003 at UTC 01h10 (Sunyaev et al. 2003), when 
the source was bright for $\sim$ 2 hours
and then faded below the detection threshold. 
About 5 hours later the source turned on again (Grebenev et al. 2003) and underwent another fast outburst which  lasted $\sim$ 8 hours and 
was characterized by 2 peaks having flux values of  $\sim$ 60 and  $\sim$ 80 mCrab , respectively (25--50 keV).
INTEGRAL again detected the source in outburst on 8 March 2004 (Grebenev et al. 2004); the duration was $\sim$ 10 hours. 

Subsequent to the INTEGRAL discovery of the source, in't Zand et al. (2004) analyzed the BeppoSAX WFC data archive revealing 
that IGR~J17544$-$2619 was already detected in outburst five times  
from 1996 to 2000. The durations vary between 10 minutes 
and 8 hours while the peak fluxes were measured to be between 100 and 200 mCrab (2--28 keV).

Three XMM observations of IGR~J17544$-$2619 (Gonzalez-Riestra et al. 2004) showed that its flux varies strongly on very short timescales (few minutes). 
Furthermore they provide an upper limit to the flux in quiescence of  $F_{0.5{-}10~\rm keV}$ $\leq$  5$\times$10$^{-14}$ erg cm$^{-2}$ s$^{-1}$ while 
the flux in outburst was 7.5$\times$10$^{-11}$ erg cm$^{-2}$ s$^{-1}$.
Chandra observed the source both in quiescence and outburst on 3 July 2004 (in't Zand 2005), the luminosity (0.5--10 keV) in quiescence 
is $\sim$ 5$\times$10$^{32}$ erg  s$^{-1}$ while  the spectrum during the outburst is hard and moderately absorbed. in't Zand (2005) suggests that the accretor 
is a neutron star. 

The optical counterpart of IGR~J17544$-$2619 has been identified thanks to the XMM and Chandra accurate positions as a blue O9Ib supergiant located 
at $\sim$ 3 kpc (Pellizza et al. 2006, quoted in in't Zand 2005 and Negueruela et al. 2005a).
\subsubsection{Analysis of IBIS/ISGRI observations and results}
Table 2 lists all IBIS detections of outbursts from IGR~J17544$-$2619.
Here we present 2 newly discovered fast X-ray outbursts (No. 4 and 5 in Table 2); moreover we show for the first time detailed ISGRI light curves
of two fast X-ray outbursts (No. 1 and 3 in Table 2) previously reported in the literature but not studied in detail.

Figure 4 shows the ISGRI light curve (20--40 keV) of the outburst which occurred on 17 September 2003 at UTC 01h10.   
The  fast transient nature of the source is clearly evident with two strong flares characterized by a very 
quick rise and decay (few tens of minutes). The peak-flux and the luminosity (20--40 keV) are $\sim$ 400 mCrab 
and $\sim$ 3.2$\times$10$^{36}$ erg s$^{-1}$ respectively.

Figure 5 shows the ISGRI light curve (20--60 keV) of the outburst which occurred on 8 March 2004. Initially  it is
characterized by several very quick flares (few minutes timescales), 
then suddenly the source flares up in $\sim$10 minutes to a 20--60 keV peak-flux of $\sim$ 240 mCrab. Subsequently the flux drops to
a low level in $\sim$ 20 minutes. 
We extracted a spectrum of this strong flare (20--60 keV) which is best fitted by a  thermal bremsstrahlung model (kT=9.5$\pm$0.9 keV, $\chi^{2}_{\nu}$=1.24, d.o.f. 14).
However a reasonable fit is also provided by a black body model (kT=4.4$\pm$0.25 keV, $\chi^{2}_{\nu}$=1.4, d.o.f. 14). 

\clearpage
\begin{table*}[t!]
\begin{center}
\caption {Summary of ISGRI observations of outbursts of IGR~J17544$-$2619}
\begin{tabular}{clcccc}
\hline
\hline
No.  & Date  & duration (hours) & flux at the peak & ref\\
\hline
1  & 17 September 2003 01h10 UTC & $\sim$ 2  & $\sim$ 400 mCrab (20--40 keV)  & [1] \\
2  & 17 September 2003 06h UTC& $\sim$ 8 & $\sim$ 80 mCrab (25--50 keV) & [2] \\  
3  & 8 March 2004 &  $\sim$ 10 &  $\sim$ 240 mCrab (20--60 keV)& [3] \\
4  & 21 September 2004 &  --- &  $\sim$ 70 mCrab (20--40 keV)  & $\star$  &   \\
5  & 12 March 2005 &  $\sim$ 0.5 &  $\sim$ 150 mCrab (20--30 keV)  &  $\star$ \\
\hline
\hline
\end{tabular}
\end{center}
[1] Sunyaev et al. 2003 [2] Grebenev et al. 2003  [3] Grebenev et al. 2004 \\ $\star$ = this paper \\
\end{table*} 
\clearpage
\begin{figure}[t!]
\plotone{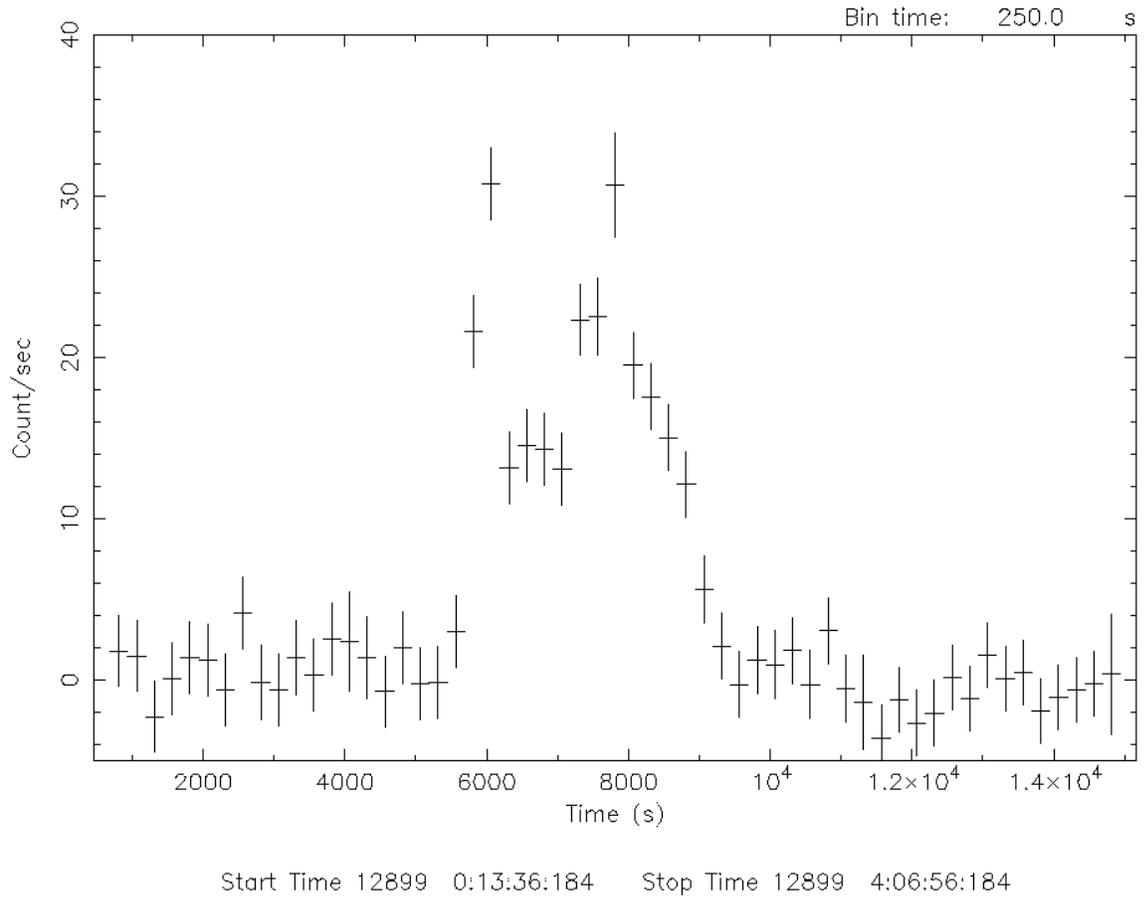}
\caption{The ISGRI light curve (20--40 keV) of the IBIS detection of IGR~J17544$-$2619 on 17 September 2003 at UTC 01h10.}
\end{figure}
\clearpage
\begin{figure}[t!]
\plotone{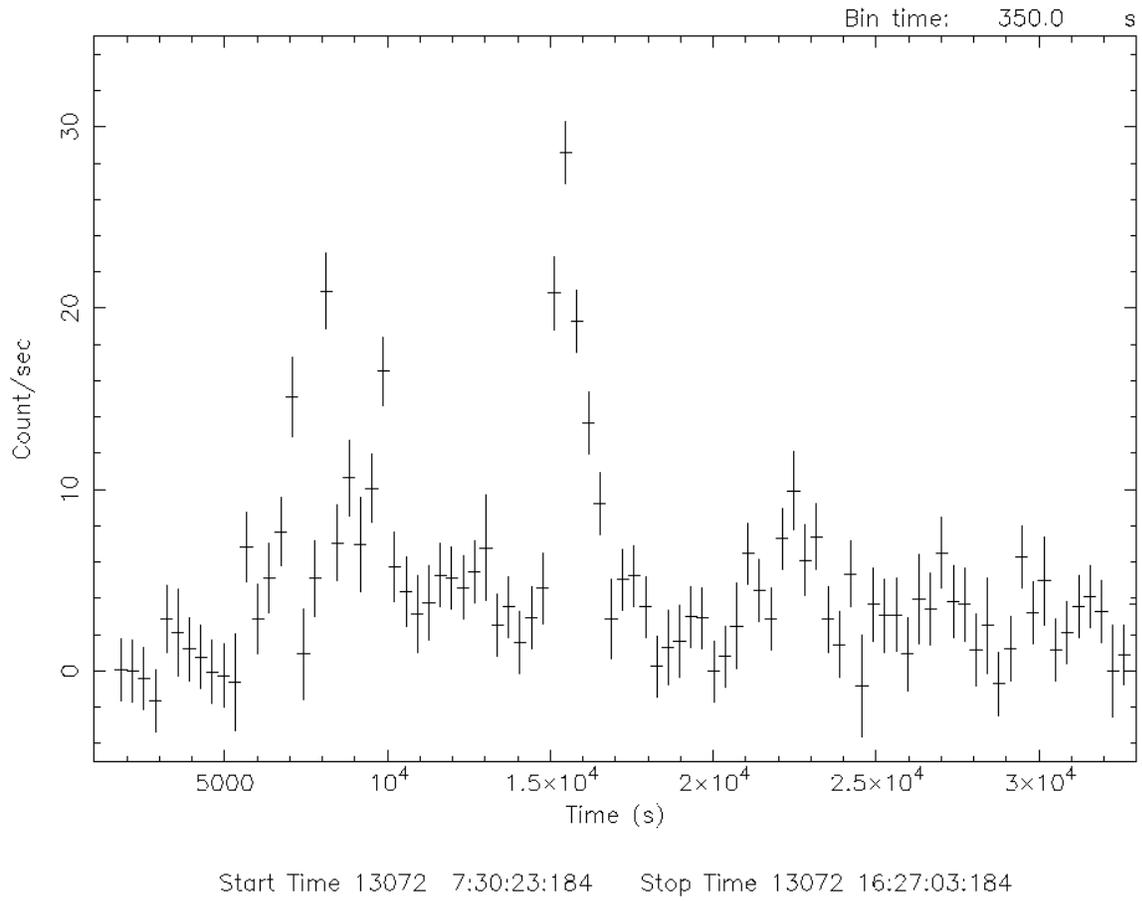}
\caption{The ISGRI light curve (20--60 keV) of the IBIS detection of IGR~J17544$-$2619 on 8 March 2004.}
\end{figure}

\clearpage
As for the two newly discovered fast X-ray outbursts, the first 
occurred on 21 September 2004 and its 20--40 keV light curve is shown in Figure 6.
Unfortunately the light curve is truncated at the beginning because the source was outside the IBIS field of view, 
so we do not constrain the duration of the outburst.  The  peak-flux (20--40 keV) 
was $\sim$70 mCrab.  

The second outburst occurred on 12 March 2005 and it was detected in only one ScW. As we can see from the 20--30 keV ISGRI light curve (Figure 7),
the data are of poorer quality and don not allow any fast variability to be seen.
The duration  and the peak-flux were $\sim$ 30 minutes and $\sim$ 150 mCrab (20--30 keV) respectively. 
Although the source was detected in only one ScW, we have extracted a  spectrum during the flare.
The best fit model (20--40 keV) is provided by a black body  (kT=2.9$\pm$0.4 keV, $\chi^{2}_{\nu}$=1.007, d.o.f. 8) (see Figure 8).
However, even  an optically thin thermal bremsstrahlung model provides a reasonable  fit  (kT=5$\pm$1 keV, $\chi^{2}_{\nu}$=0.8, d.o.f. 8);
this type of spectrum in the hard X-ray band is typical of binary systems with neutron stars.

\clearpage
\begin{figure}[t!]
\plotone{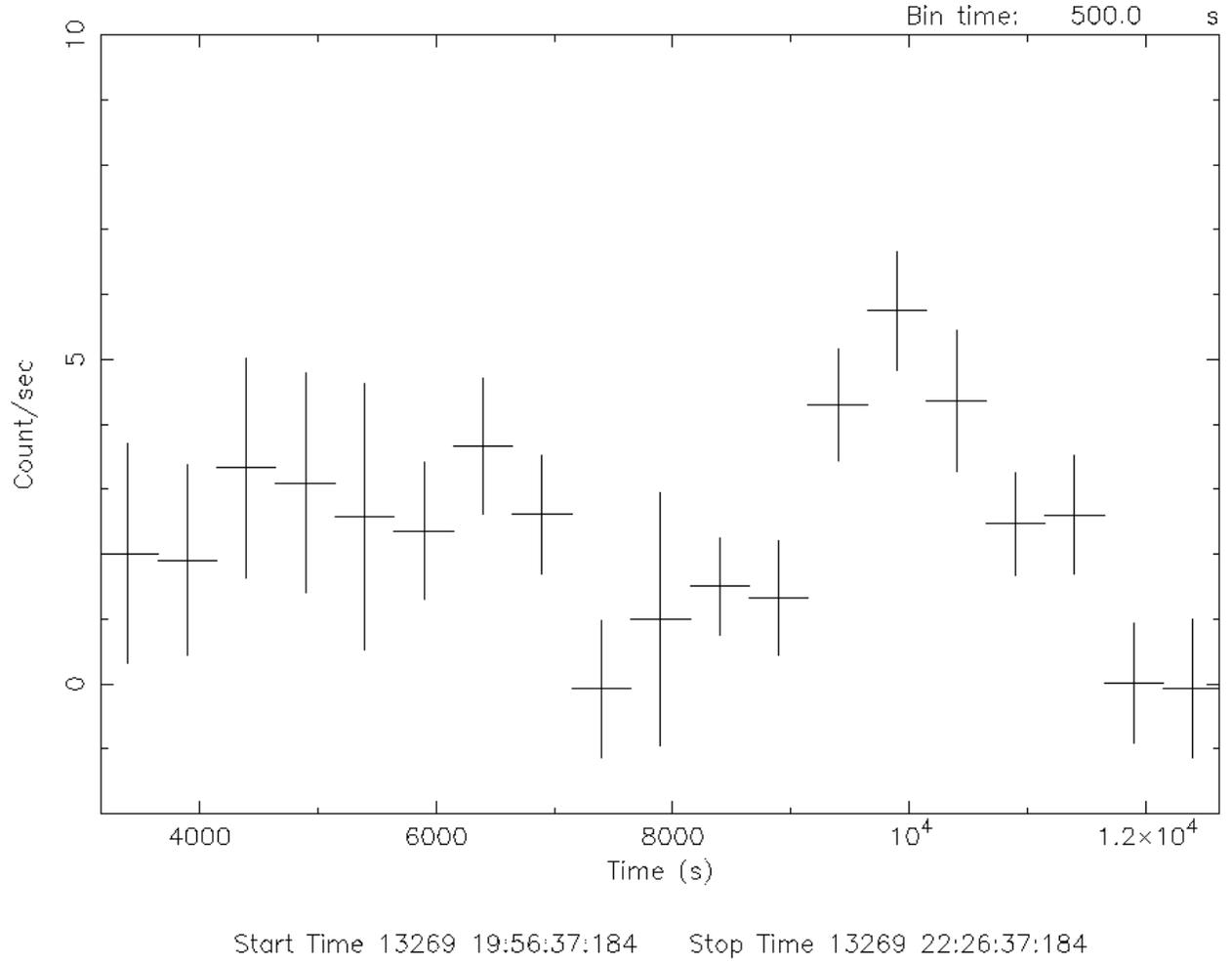}
\caption{The ISGRI light curve (20--40 keV) of a newly discovered outburst of IGR~J17544$-$2619 detected on  21 September 2004.}
\end{figure}
\clearpage
\begin{figure}[t!]
\plotone{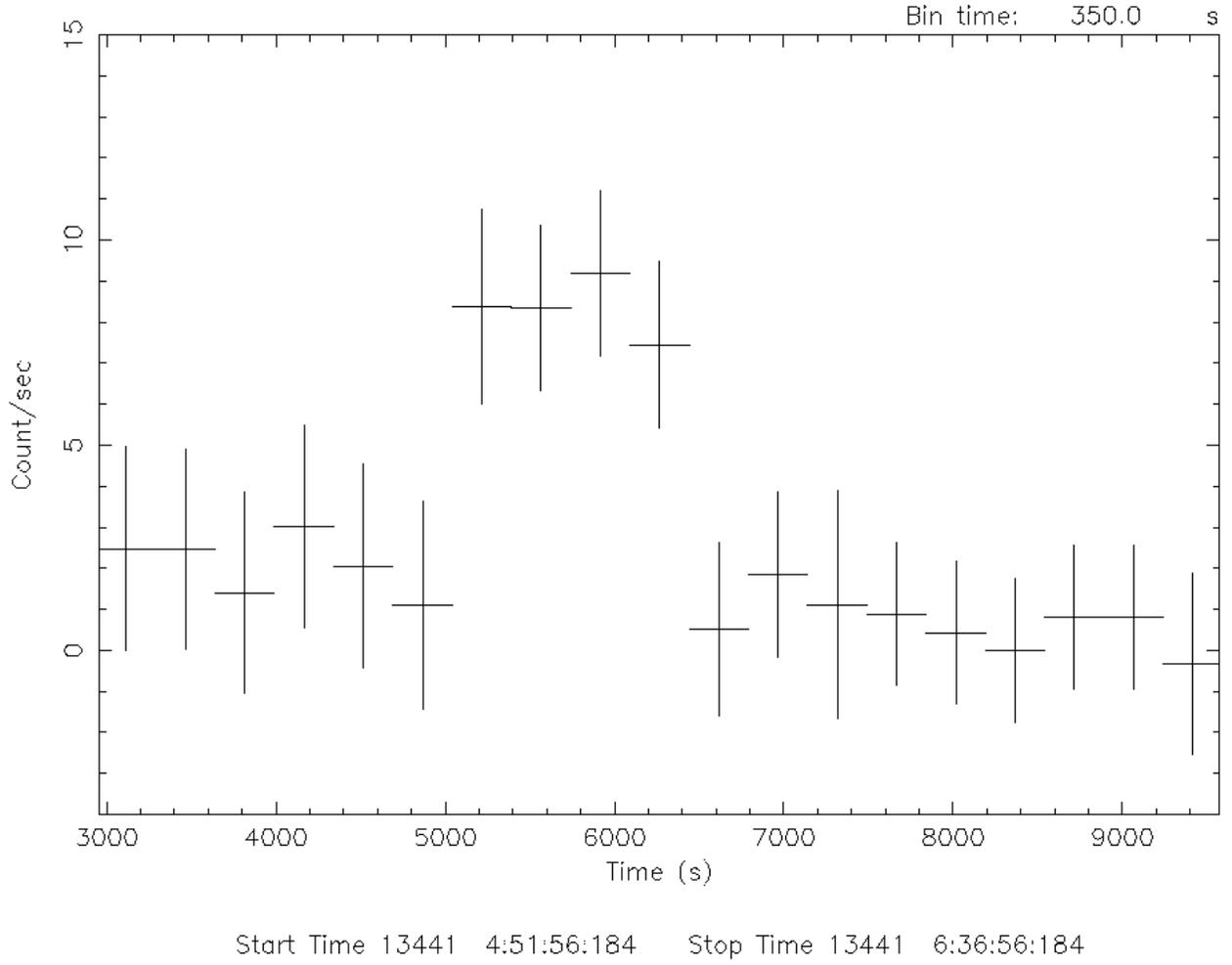}
\caption{the ISGRI light curve (20--30 keV) of a newly discovered outburst of IGR~J17544$-$2619 detected on  12 March 2005.}
\end{figure}
\begin{figure}[t!]
\plotone{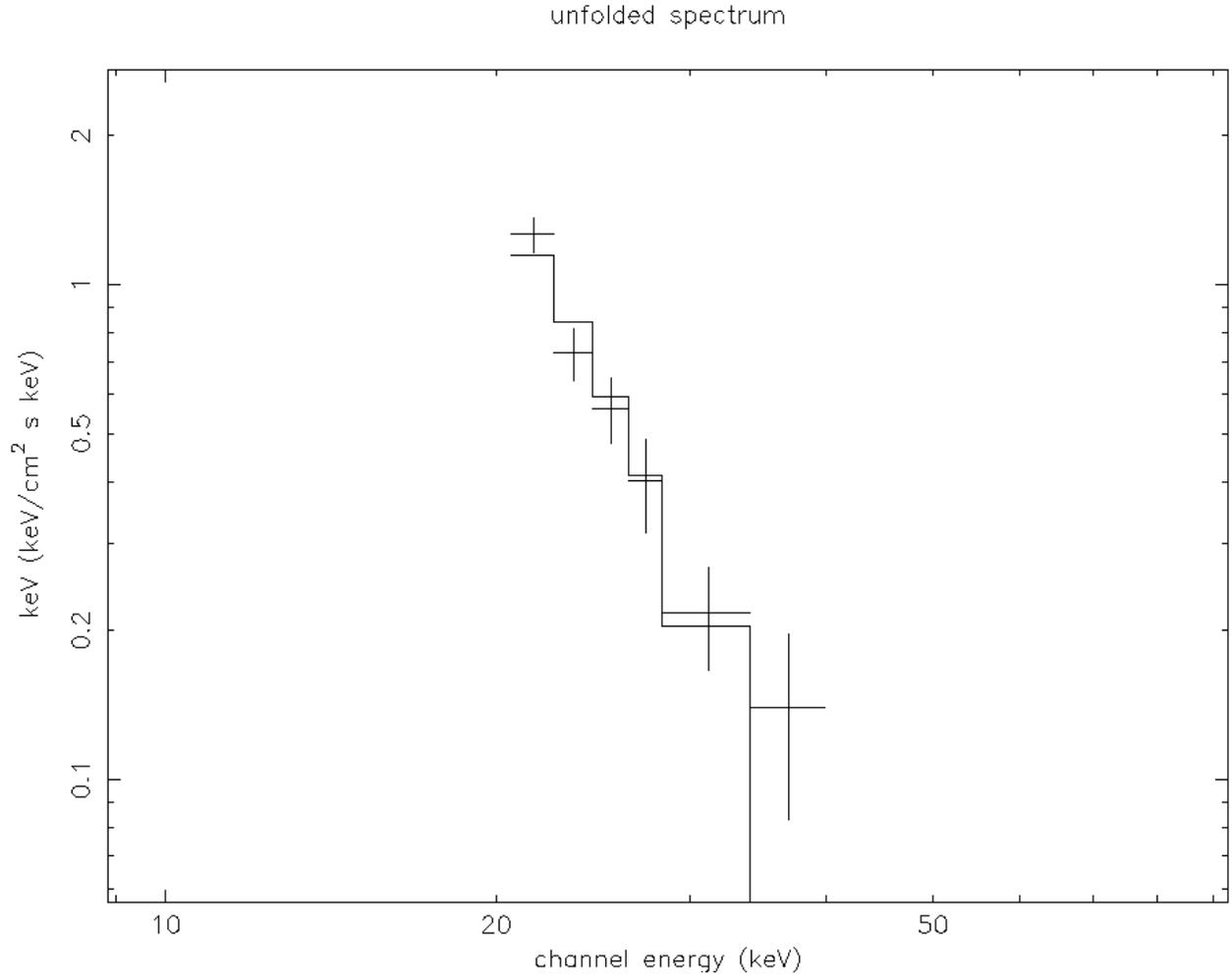}
\caption{The unfolded black body spectrum (20--40 keV) of  IGR~J17544$-$2619 during the outburst that occurred on  12 March 2005.}
\end{figure}
\clearpage

\subsection{IGR~J18410$-$0535=AX~J1841.0$-$0536}

\subsubsection{Archival X-ray observations of the source}
IGR~J18410$-$0535 was discovered during a GCDE on 8 October 2004 as it was undergoing an outburst detected between 20--60 keV (Rodriguez et al. 2004).
The nominal position of IGR~J18410$-$0535  is in agreement with the Chandra location (Halpern \& Gotthelf 2004) of the ASCA 4.74 s transient pulsar 
AX~J1841.0$-$0536  (Bamba et al. 2001), indicating that they are the same object. AX~J1841.0$-$0536   was detected  by ASCA as a fast X-ray transient 
source in April 1994 and October 1999, in both cases a noticeable  feature was multiple flares with a fast rise time of a few hours.
The accurate Chandra position  
permitted the identification  of the optical counterpart of  AX~J1841.0$-$0536 with a luminous supergiant star (Negueruela et al. 2005a),
while the spectrum is well fitted with an absorbed  power law ($\Gamma$=1.35, N$_{H}$=6$\times$10$^{22}$ cm$^{-2}$) (Halpern et al. 2004).

\subsubsection{Analysis of IBIS/ISGRI observations and results}
Table 3 lists all IBIS detections of fast X-ray outbursts from  IGR~J18410$-$0535.
Here we present 2 newly discovered fast X-ray outbursts (Nos. 1 and 2 in Table 3).

The first outburst occurred on 
24 April 2003  and  was detected in the energy band  20--80 keV for $\sim$ 7 hours. 
As we can note in Figure 9, the light curve shows a gradual increase,
then suddenly the source flares up 
to a peak very quickly ($\sim$ 20 minutes) and subsequently it drops off with a similar timescale as the rise. The peak-flux  is 
$\sim$ 120 mCrab (20--80 keV).
A spectrum extracted during the outburst is best fitted by a black body model (kT=8$\pm$0.9 keV, $\chi^{2}_{\nu}$=1.1, d.o.f. 19).

A second, shorter,  outburst  ($\sim$ 1.1 hours)  occurred on 6 May 2003. Figure 10 shows its 20--80 keV ISGRI light curve. Initially the flux is 
consistent with zero, then suddenly the source turns on, flares up and quickly reaches a peak (rise time $\sim$ 13 minutes). 
The 20--80 keV peak-flux is equal to $\sim$ 120 mCrab.
The  flux then  drops to a low level in $\sim$ 30 minutes, subsequently undergoing a small flare  after which it turned off. 
We extracted a spectrum during this outburst, the best fit (20--80 keV)  is provided
by a black body model (kT=9$\pm$0.9 keV, $\chi^{2}_{\nu}$=1.45, d.o.f. 19).
It is worth noting that both outbursts are very hard and are  detected up to 80 keV. 
To date IGR~J18410$-$0535=AX~J1841.0$-$0536 is the only SFXT to show a hard energy tail.
Our ISGRI analysis of the 2 newly discovered outbursts provided a refined position (J2000, RA=18$^{h}$ 40$^{m}$ 57.6$^{s}$, DEC=-05$^{\circ}$ 35$^{'}$ 38$^{''}$, 
1.2$^{'}$  error radius) which is located $\sim$ 40$^{''}$ from the Chandra position of AX~J1841.0$-$0536.
\clearpage
\begin{table*}[t!]
\begin{center}
\caption {Summary of ISGRI observations of outbursts of IGR~J18410$-$0535=AX~J1841.0$-$0536}
\begin{tabular}{clcccc}
\hline
\hline
No.  & Date  & duration (hours) & flux at the peak & ref\\
\hline
1  & 24 April 2003 & $\sim$ 7 & $\sim$ 120 mCrab (20--80 keV) &   $\star$  &   \\
2  & 6 May 2003 & $\sim$ 1.1 & $\sim$ 120 mCrab (20--80 keV) &   $\star$  &   \\
3  & 8 October 2004 &  -- &  -- &  [1] \\
\hline
\hline
\end{tabular}
\end{center}
[1] Rodriguez et al. 2004 \\ $\star$ = this paper \\
\end{table*} 
\clearpage
\begin{figure}[t!]
\plotone{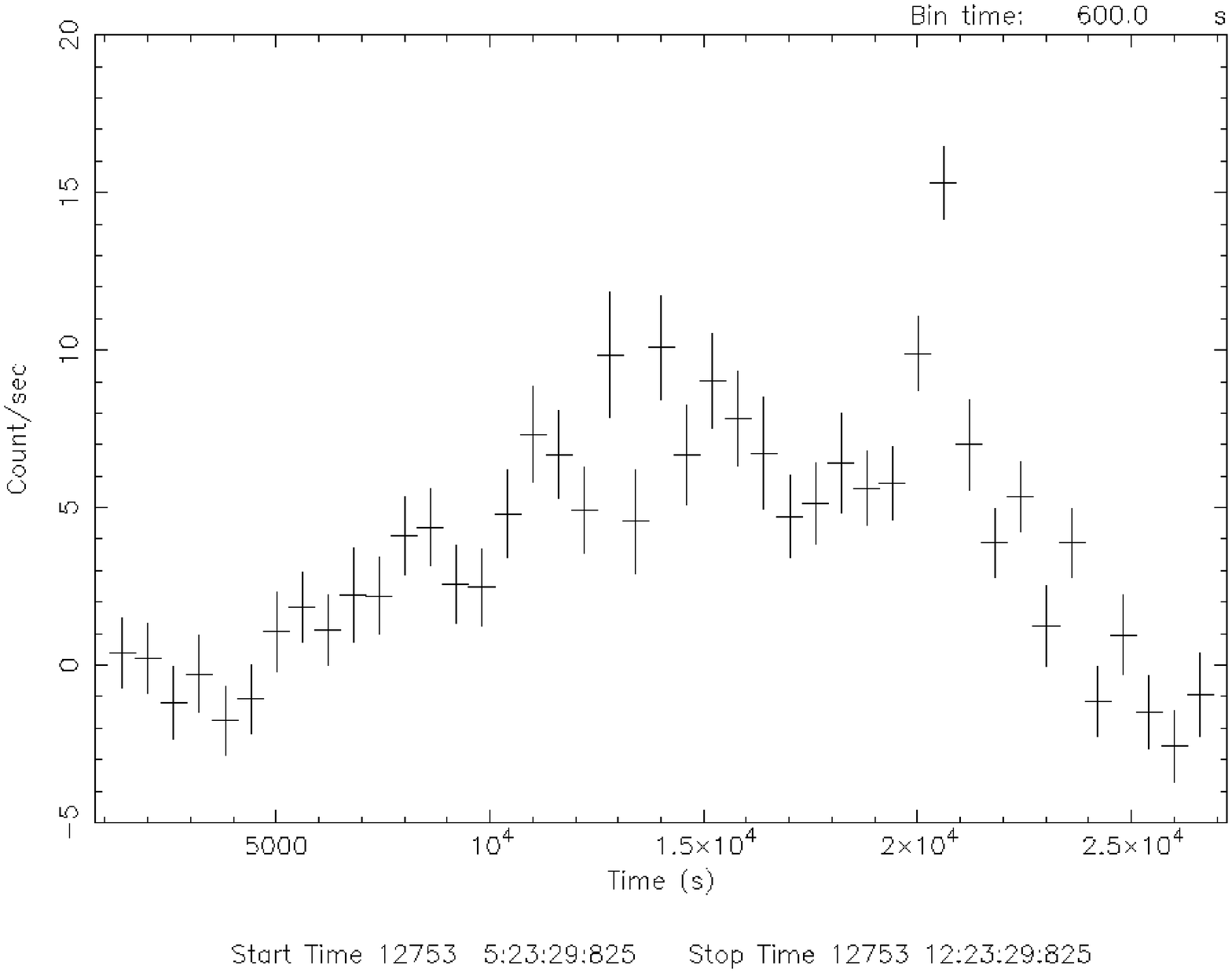}
\caption{The ISGRI light curve (20--80 keV) of a newly discovered outburst of  IGR~J18410$-$0535=AX~J1841.0$-$0536 detected on 24 April 2003.}
\end{figure}
\clearpage
\begin{figure}[t!]
\plotone{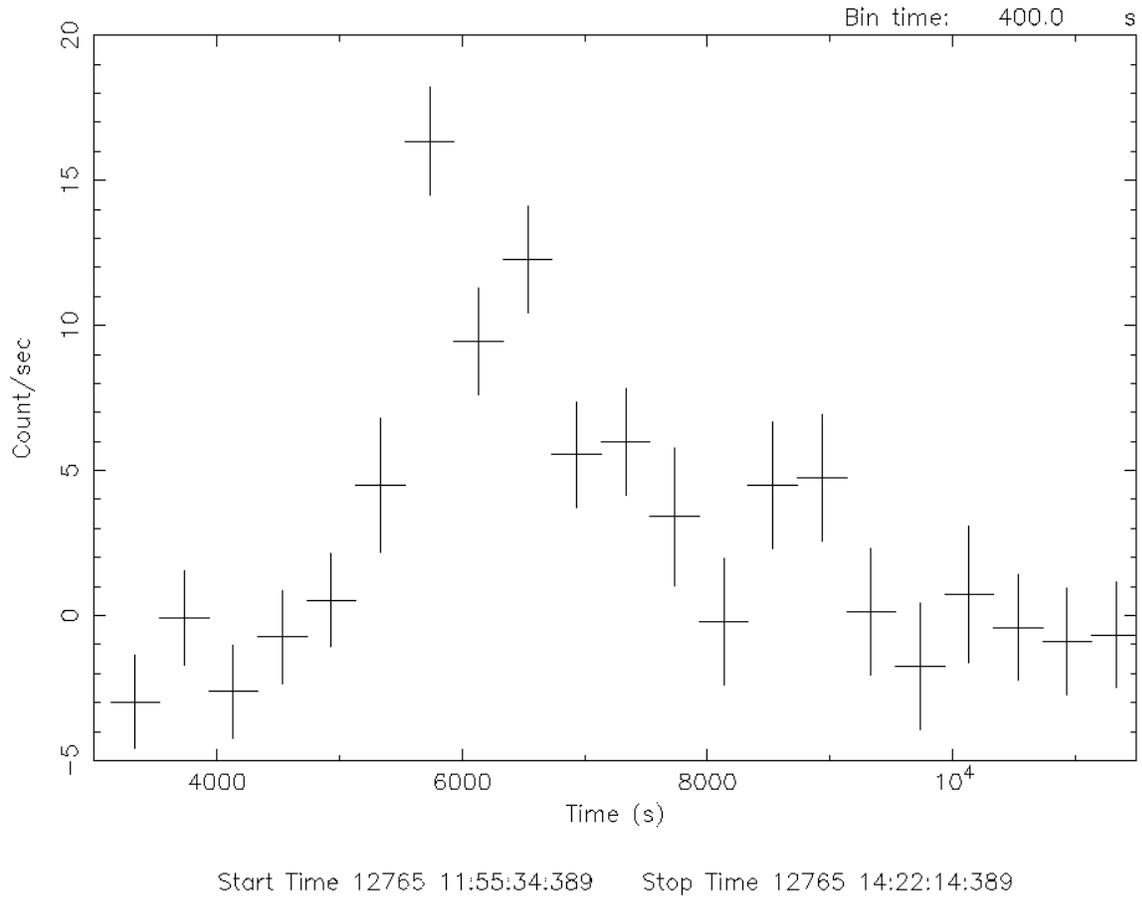}
\caption{The ISGRI light curve (20--80 keV) of a newly discovered outburst of  IGR~J18410$-$0535=AX~J1841.0$-$0536 detected on 6 May 2003.}
\end{figure}

\clearpage

\subsection{IGR~J11215$-$5952}
\subsubsection{Archival X-ray observations of the source}
This transient source was discovered on 22 April 2005 at 06:04:25 UTC during the decaying phase of an outburst  (Lubinski et al. 2005).
The reported 20--60 keV flux was 75 mCrab declining to 44 mCrab in 40 minutes. 
To date this is the only outburst to be reported in  literature.

The B1Ia-type supergiant HD 306414 (Vijapurkar \& Drilling 1993), located 16$^{''}$ from the nominal
position of the source (Negueruela et al. 2005c), is the brightest optical object in the 3$'$ ISGRI error circle.
Based on its photometric properties, Negueruela et al. (2005c) suggested that  HD 306414 is the optical
counterpart of IGR~J11215$-$5952. Masetti et al. (2005) performed an optical follow-up observation of  HD 306414 and the spectrum indicates
that it is an early B-type luminous star in agreement with the  B1Ia spectral type classification of Vijapurkar \& Drilling (1993).   
Its distance ($\sim$ 6.2 kpc) is compatible with a location in the far end of the Carina Arm.
\subsubsection{Analysis of IBIS/ISGRI observations and results}
We report on a newly discovered fast X-ray outburst of IGR~J11215$-$5952.

The source displayed fast X-ray transient activity on 4 July 2003. Figure 11 shows its ISGRI light curve; 
the peak-flux and luminosity (20--40 keV) are $\sim$ 40 mCrab and  $\sim$ 1.4$\times$10$^{36}$ erg s$^{-1}$, the latter is  in good agreement with typical 
outburst luminosities of SFXTs.
The spectrum during the outburst activity 
can be reasonably described 
($\chi^{2}_{\nu}$=0.9, d.o.f. 14) by an  optically thin thermal bremsstrahlung model  (kT=19$^{+5}_{-3.5}$ keV).
However, a  reasonable fit was also achieved using a black body model (kT=6.2$\pm$0.6 keV, $\chi^{2}_{\nu}$=1.4, d.o.f. 14).

Our ISGRI analysis of this outburst provided a position at  
RA=11$^{h}$ 21$^{m}$ 50.16$^{s}$ DEC=-59$^{\circ}$ 52$^{'}$ 04.8$^{''}$ (J2000) with an error radius of 2.2$^{'}$, which is smaller than that reported when  
the source was discovered ($\sim$ 3$^{'}$, Lubinski et al. 2005).  Obviously, a more refined  position is needed in order  to confirm 
the association between IGR~J11215$-$5952 and  HD 306414.

\clearpage
\begin{figure}[t!]
\plotone{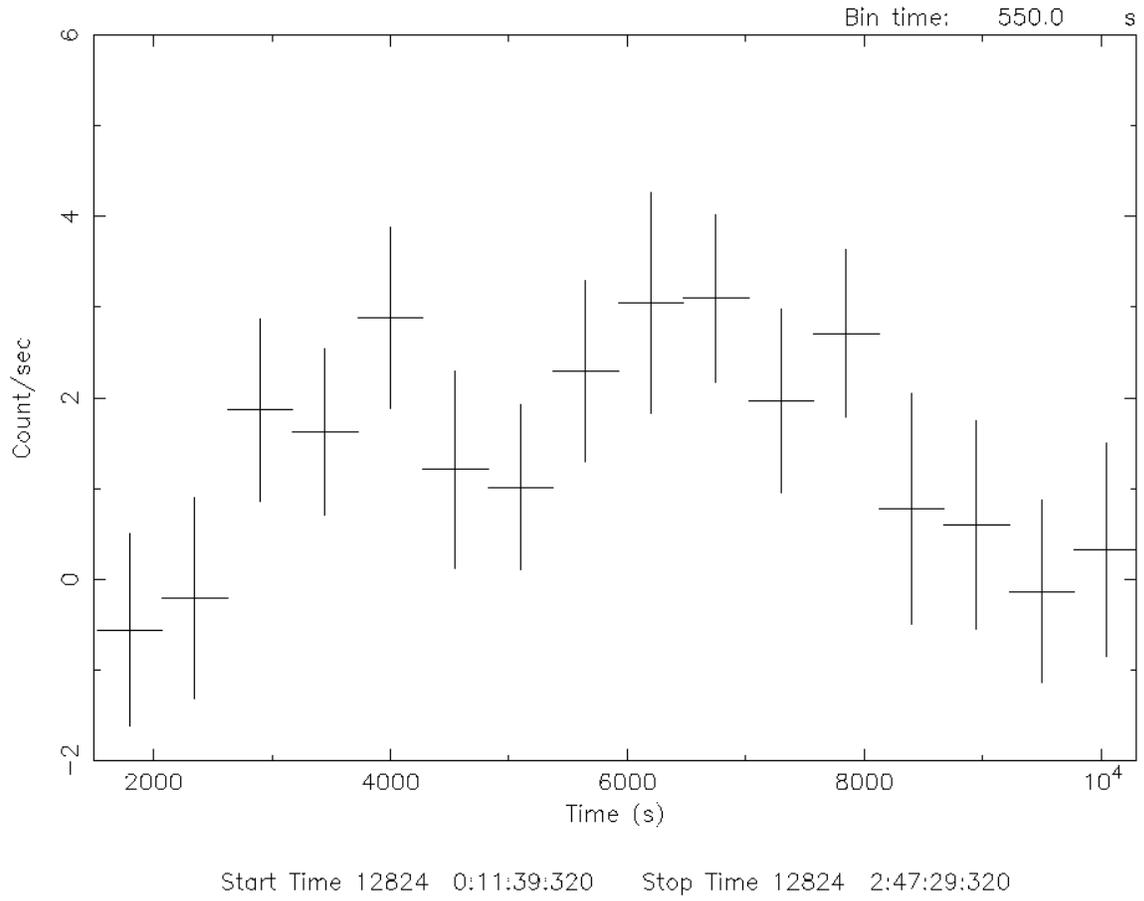}
\caption{The ISGRI light curve (20--40 keV) of a newly discovered outburst of IGR~J11215$-$5952.}
\end{figure}
\clearpage

\section{Possible candidate supergiant fast X-ray transients} 

\subsection{IGR~J17407$-$2808}
IGR~J17407$-$2808 was discovered on 9 October 2004 (Kretschmar et al. 2004) as it was undergoing strong X-ray outburst activity 
characterized by several flares  detected in the  20--60 keV band. The nominal position is (J2000) RA=17$^{h}$ 40$^{m}$ 42$^{s}$ 
DEC=-28$^{\circ}$ 08$^{'}$ 00$^{''}$ with an error circle of 2.3$^{'}$ radius.
The most energetic flare was strong enough to trigger an automatic alert message  of the INTEGRAL Burst Alert  System IBAS (Mereghetti et al. 2003)
but the position consistency with the X-ray source SBM2001 10 (50$^{''}$ angular separation between them)
and soft spectrum (Gotz et al. 2004) excluded  a gamma-ray burst origin for IGR~J17407$-$2808. 

SBM2001 10 is a faint unidentified X-ray source listed in the ROSAT catalog of sources in the Galactic Center region (Sidoli et al. 2001).
Its position is (J2000) RA=17$^{h}$ 40$^{m}$ 41.2$^{s}$ DEC=-28$^{\circ}$ 08$^{'}$ 50$^{''}$ (error circle of 16$^{''}$ radius) while the count rate is equal to 
3.73$\pm$1.2 cts ksec$^{-1}$ (0.1--2.4 keV).

We extracted a 20--60 keV ISGRI light curve (bin time of 25 seconds) from this detection (Figure 12).
Three prominent very fast flares, lasting no more than few minutes, are clearly visible. In particular the last flare is very strong,  reaching  a 
peak-flux of $\sim$ 805  mCrab or 9.5$\times$10$^{-9}$ erg cm$^{-2}$ s$^{-1}$ (20--60 keV). The rise and decay times are  $\sim$ 50 s and  $\sim$25 s respectively.
\clearpage
\begin{figure}[t!]
\plotone{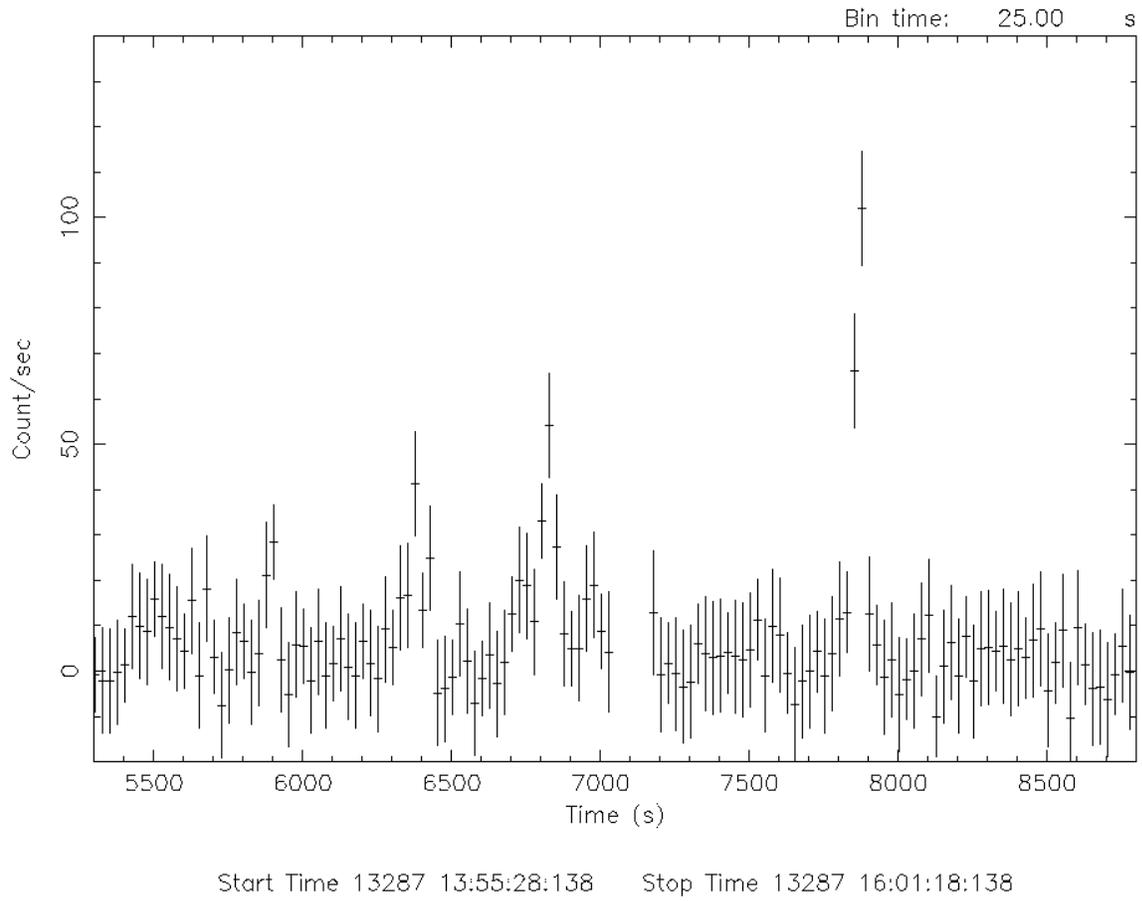}
\caption{The ISGRI light curve (20--60 keV) of the detection of IGR~J17407$-$2808 on 9 October 2004.}
\end{figure}

\clearpage
We used a Good Time Interval (GTI) analysis to
extract the spectrum of IGR~J17407$-$2808 during this fast and strong flare lasting $\sim$ 1 minute.
The 20--60 keV spectrum  is best fitted with a thermal bremsstrahlung model (kT=23$^{+7}_{-4.5}$ keV, $\chi^{2}_{\nu}$=0.78, d.o.f. 14);  
however a black body model also 
gives  a reasonable  fit (kT=7$\pm$0.7 keV, $\chi^{2}_{\nu}$=1.3, d.o.f. 14).

Our IBIS analysis of this flare provided a more accurate source position (J2000, RA=17$^{h}$ 40$^{m}$ 40.08$^{s}$ 
DEC=-28$^{\circ}$ 08$^{'}$ 24$^{''}$, error radius of 1.7$^{'}$) which is 
located 30$^{''}$ from the ROSAT source SBM2001 10. Even if the ROSAT error circle is quite small (16$''$ radius), it  could not be sufficiently small 
so as to allow an optical follow-up observation which is necessary to provide unambigous identification.  

It is worth noting that IGR~J17407$-$2808 seems to be a very peculiar fast X-ray transient source. It was detected 
only once by IBIS, and 
its outburst activity was characterized by 3 very quick and strong flares (20--60 keV, peak-flux $\sim$ 800 mCrab) with timescales 
less than a couple of minutes. This makes the outburst activity of  IGR~J17407$-$2808 significantly
shorter than typical flaring activity from SFXTs which lasts at least a few hours.
We cannot rule out the possibility that this outburst is not from a supergiant HMXB. Its timing behaviour resembles the so called 
burst-only sources detected by the WFCs on board BeppoSAX up to 30 keV (Cornelisse et al. 2004), although IGR~J17407$-$2808 was  detected by IBIS up to 60 keV. 
 
\subsection{IGR~J16418$-$4532}
Tomsick et al. (2004) reported the discovery of IGR~J16418$-$4532 during an INTEGRAL observation targeted to the black hole X-ray 
transient 4U 1630-47 and performed between 1--5 February 2003. The 20--40 keV flux was 3$\times$10$^{-11}$ erg cm$^{-2}$ s$^{-1}$. To date this 
is the only outburst to be reported from this source.

We report on a newly discovered fast X-ray outburst of IGR~J16418$-$4532 
which unveils for the first time its likely fast X-ray transient nature.
It was detected by IBIS in only 2 consecutive ScWs (20--30 keV)
on 26 February 2004 (duration of  $\sim$ 1 hour). Figure 13 shows 
the 20--30 keV ISGRI light curve, while the peak-flux was $\sim$ 80 mCrab.    
\clearpage
\begin{figure}[t!]
\plotone{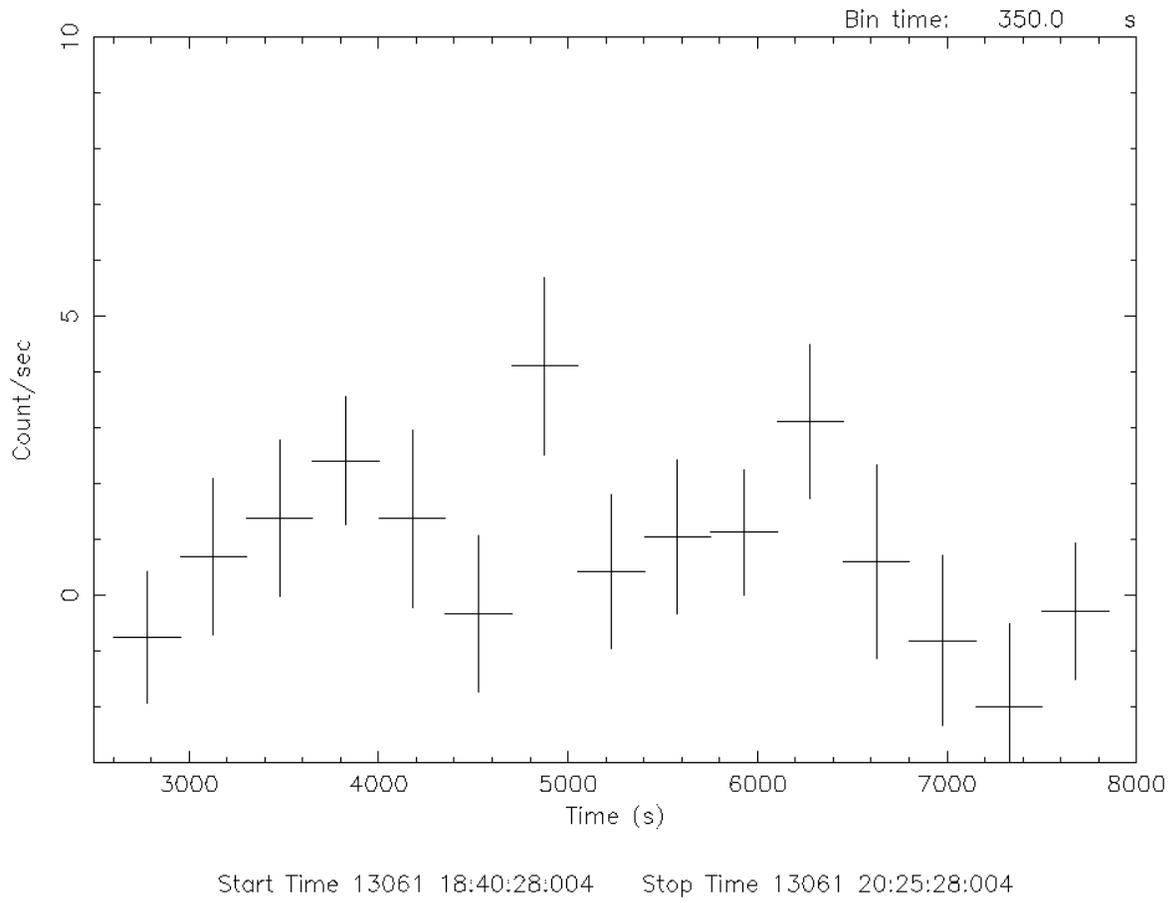}
\caption{The ISGRI light curve (20--30 keV) of a newly discovered outburst of IGR~J16418$-$4532.}
\end{figure}
\clearpage

\subsection{IGR~J16479$-$4514}
\subsubsection{Archival X-ray observations of the source}
IGR~J16479$-$4514 was discovered with IBIS during observations of the Galactic Center region performed on 8--9 August 2003 
(Molkov et al. 2003). The average fluxes were  12 mCrab and 8 mCrab in the energy bands 18--25~keV and 25--50~keV, respectively. 
During observations performed on August 10, the source showed outburst activity where the flux increased by a factor $\sim$2 
in the same energy bands (Molkov et al. 2003). These outbursts were reported as an average detection, 
hence the fast X-ray transient nature of the source was not reported. We have performed an analysis at the ScW level
 and we show for the first time the ISGRI light curve of the source
on 10 August 2003, where the fast X-ray transient behaviour is clearly evident (Figure 14). 
The duration of the outburst was $\sim$ 3.5 hours with a 20--30 keV peak-flux 
of $\sim$ 150 mCrab.
This kind of  fast transient behaviour is confirmed by the detections of 4 more fast X-ray  outbursts from IGR~J16479$-$4514
reported by Sguera et al. (2005). The durations vary from $\sim$ 30 minutes to $\sim$ 3 hours; in particular the shortest outburst was 
also the most energetic,  reaching a 20--30 keV peak-flux of 850 mCrab (Sguera et al. 2005). The energetic and temporal behavior of this outburst 
is reminiscent of a type I X-ray burst, however this kind of explanation is inconclusive since the statistics are insufficient to find evidence of
any spectral softening (Sguera et al. 2005).

Lutovinov et al. (2005a) published a broad band energy spectrum of IGR~J16479$-$4514 (1--100~keV) described by a simple power law
model modified by a cutoff at high energies and photoabsorption at soft X-rays (N$_{H}$=1.2$\times$10$^{23}$ cm$^{-2}$) exceeding the galactic value along the line of sight.
Based on these spectral characteristics, they suggested that it is a neutron star binary system with a high mass companion.

On 30 August 2005, the Swift X-ray Telescope (XRT) detected flaring activity from IGR~J16479$-$4514  
(Kennea et al. 2005) consisting of 2 fast flares (15--50 keV).
The spectrum of the source was fitted by an absorbed power law ($\Gamma$=1.1) with a 0.5--10 keV flux of  3.8$\times$10$^{-11}$ erg cm$^{-2}$ s$^{-1}$.
The Swift XRT observation provided a very accurate source position  (J2000, RA=16$^{h}$ 48$^{m}$ 07$^{s}$, DEC=-45$^{\circ}$ 12$^{'}$ 05.8$^{''}$)
with an error circle of 6$^{''}$ radius inside which the Swift optical telescope detected a faint object 
(J2000, RA=16$^{h}$ 48$^{m}$ 06.8$^{s}$, DEC=-45$^{\circ}$ 12$^{'}$ 08$^{''}$)
 in the V band with a magnitude of 20.4$\pm$0.4. This optical source is catalogued
in the USNO--B1 catalog with  I and R1 magnitudes of 16.7 and 18.36, respectively. It is the only USNO--B1 object
inside the Swift error box, and an optical follow-up observation is necessary to unveil its nature.
\clearpage
\begin{figure}[t!]
\plotone{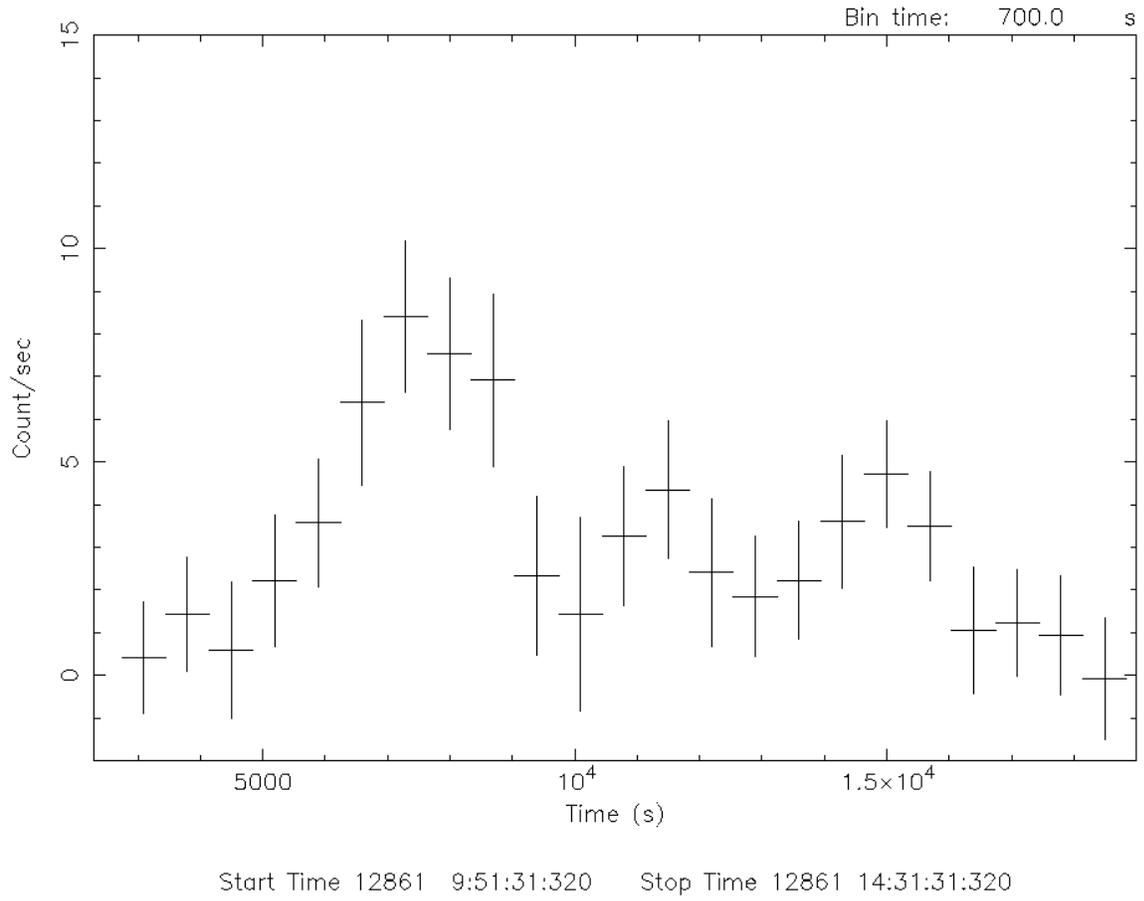}
\caption{The ISGRI light curve (20--30 keV) of the detection of  IGR~J16479$-$4514 on 10 August 2003.}
\end{figure}
\clearpage
\subsubsection{Analysis of IBIS/ISGRI observations and results}
Table 4 lists all IBIS detections of outbursts from IGR~J16479$-$4514.
Here we report on 3 newly discovered fast X-ray outbursts (Nos. 6, 7 and 8 in Table 4).              .

The first occurred on 7 September 2004 (Figure 15) with  a duration of $\sim$ 2.2 hours and a peak-flux (20--60 keV) of $\sim$ 125 mCrab. It was characterized by 2 very quick
flares, the first  with both a fast rise and decay timescale ($\sim$ 10 minutes) while the second one has the same fast rise but a slower exponential 
decay.
\begin{table*}[t!]
\begin{center}
\caption {Summary of ISGRI observations of outbursts of IGR~J16479$-$4514}
\begin{tabular}{clcccc}
\hline
\hline
No.  & Date  & duration (hours) & flux at the peak & ref\\
\hline
1  & 5 March 2003 & $\sim$ 3.5  & $\sim$ 850 mCrab (20--30 keV)  & [1] \\
2  & 28 March 2003 & $\sim$ 1.5 & $\sim$ 40 mCrab$\dagger$ & [1] \\  
3  & 21 April 2003 &  $\sim$ 0.5 &  $\sim$ 160 mCrab$\dagger$ &  [1] \\
4  & 8--10 August 2003 & -- & -- & [2] & \\
5 & 14 August 2003 & $\sim$ 2 &  $\sim$ 44 mCrab$\dagger$ & [1] \\
6  & 7 September 2004 &  $\sim$2 &  $\sim$ 125 mCrab (20--60 keV)  & $\star$  &   \\
7  & 16 September 2004 &  $\sim$ 2.5 &  $\sim$ 120 mCrab (20--60 keV)  &  $\star$ \\
8  & 4 April 2005 &  $\sim$ 2.5 &  $\sim$ 60 mCrab (20--60 keV)  &  $\star$ \\
\hline
\hline
\end{tabular}
\end{center}
[1] Sguera et al. 2005 [2] Molkov et al. 2003  \\ $\star$ = this paper \\  $\dagger$ = Average flux (20--30 keV) during the outburst
\end{table*} 

\clearpage
\begin{figure}[t!]
\plotone{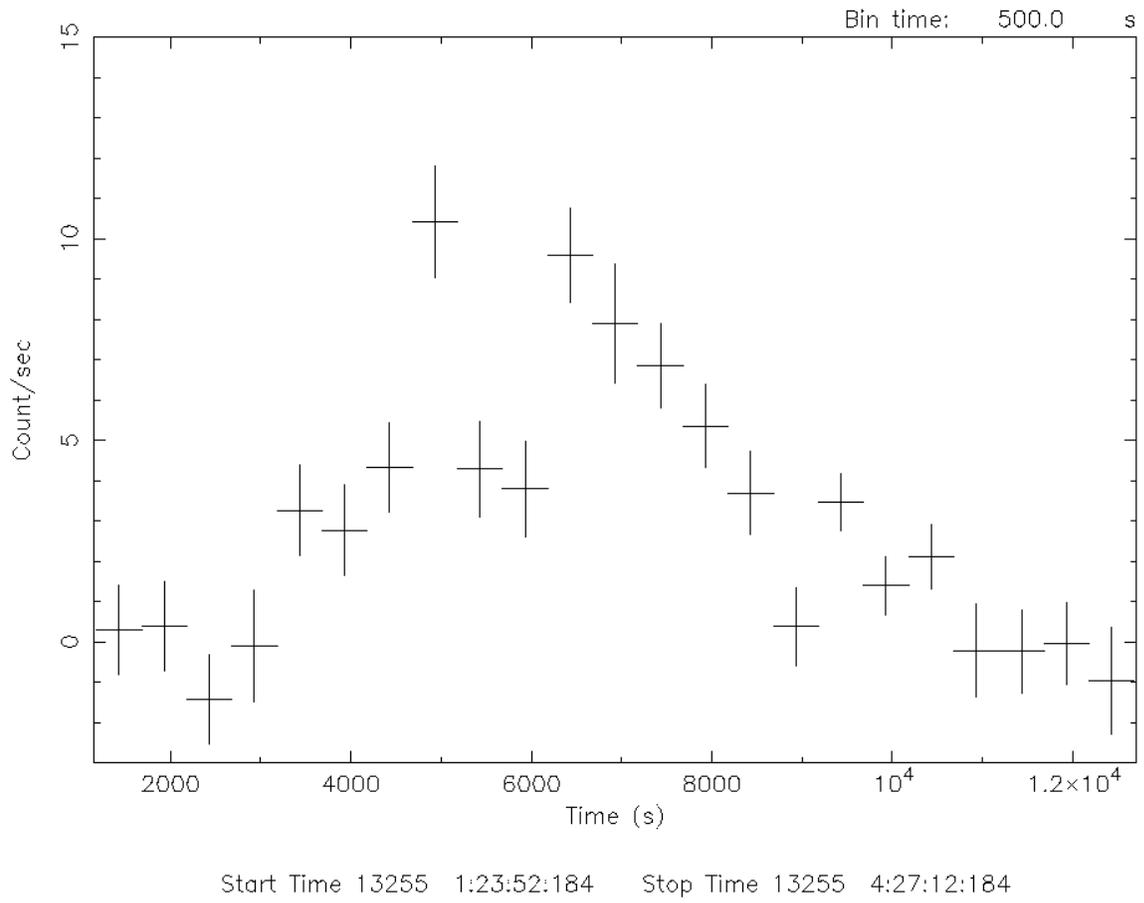}
\caption{The ISGRI light curve (20--60 keV) of a newly discovered outburst of IGR~J16479$-$4514 on 7 September 2004.}
\end{figure}
\clearpage
The second outburst was detected $\sim$ 9 days later on 16 September 2004 (Figure 16) and it is characterized by one single strong flare with similar timescales during
the rise and the decay. The duration and peak-flux are $\sim$ 2.5 hours and $\sim$ 120 mCrab (20--60 keV) respectively.
An ISGRI spectrum extracted during this outburst is equally well fitted (20--60 keV) by a black body model (kT=7.4$\pm$0.5 keV, $\chi^{2}_{\nu}$=0.95, d.o.f. 14)
or by a simple power law ($\Gamma$=2.6$\pm$0.2, $\chi^{2}_{\nu}$=1.06, d.o.f. 14).
\begin{figure}[t!]
\plotone{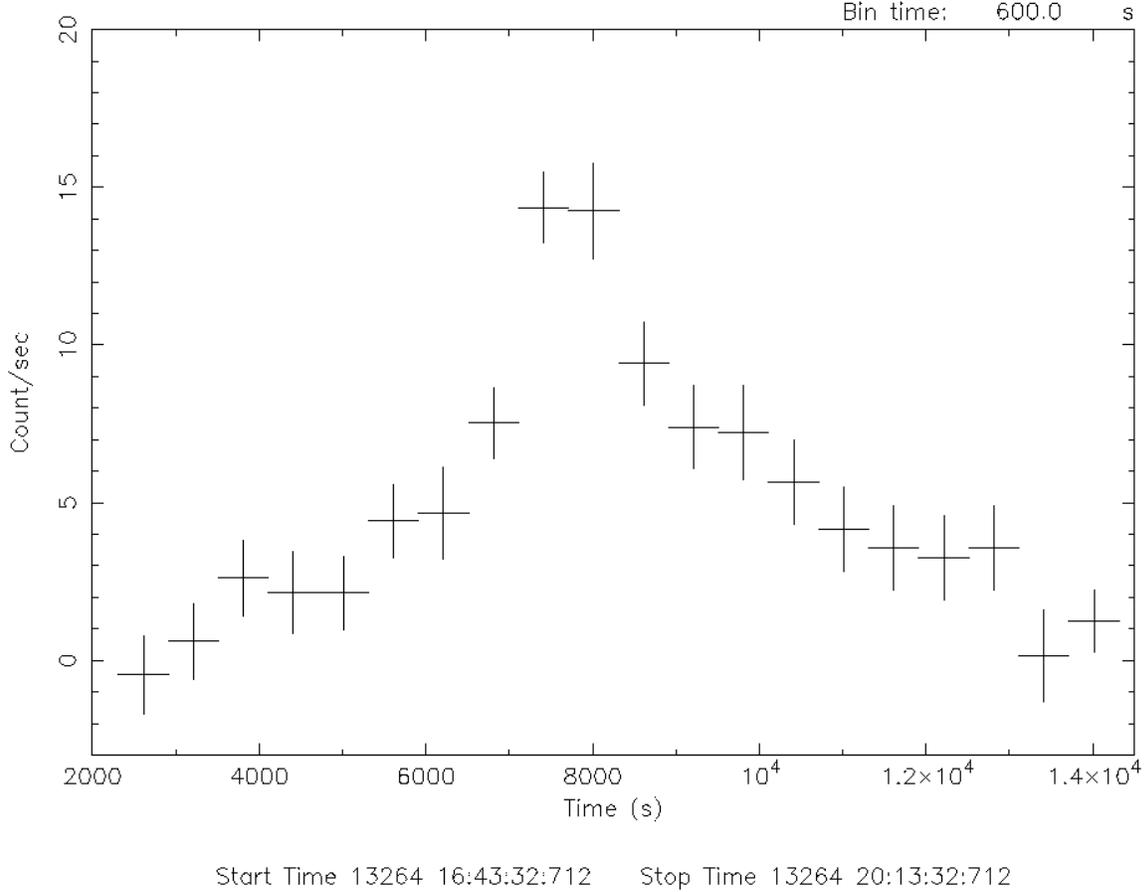}
\caption{The ISGRI light curve (20--60 keV) of a newly discovered outburst of IGR~J16479$-$4514  on 16 September 2004.}
\end{figure}
\clearpage

The last outburst occurred on 4 April 2005 (Figure 17), with  duration and peak-flux of $\sim$ 2.5 hours and $\sim$ 60 mCrab (20--60 keV) respectively.\\
The three newly discovered outbursts, together with  those previously  detected by INTEGRAL (Sguera et al. 2005), indicate a fast transient 
nature which strongly resembles that of known supergiant fast X-ray transients, therefore IGR~J16479$-$4514 can be considered a  candidate. 
\begin{figure}[t!]
\plotone{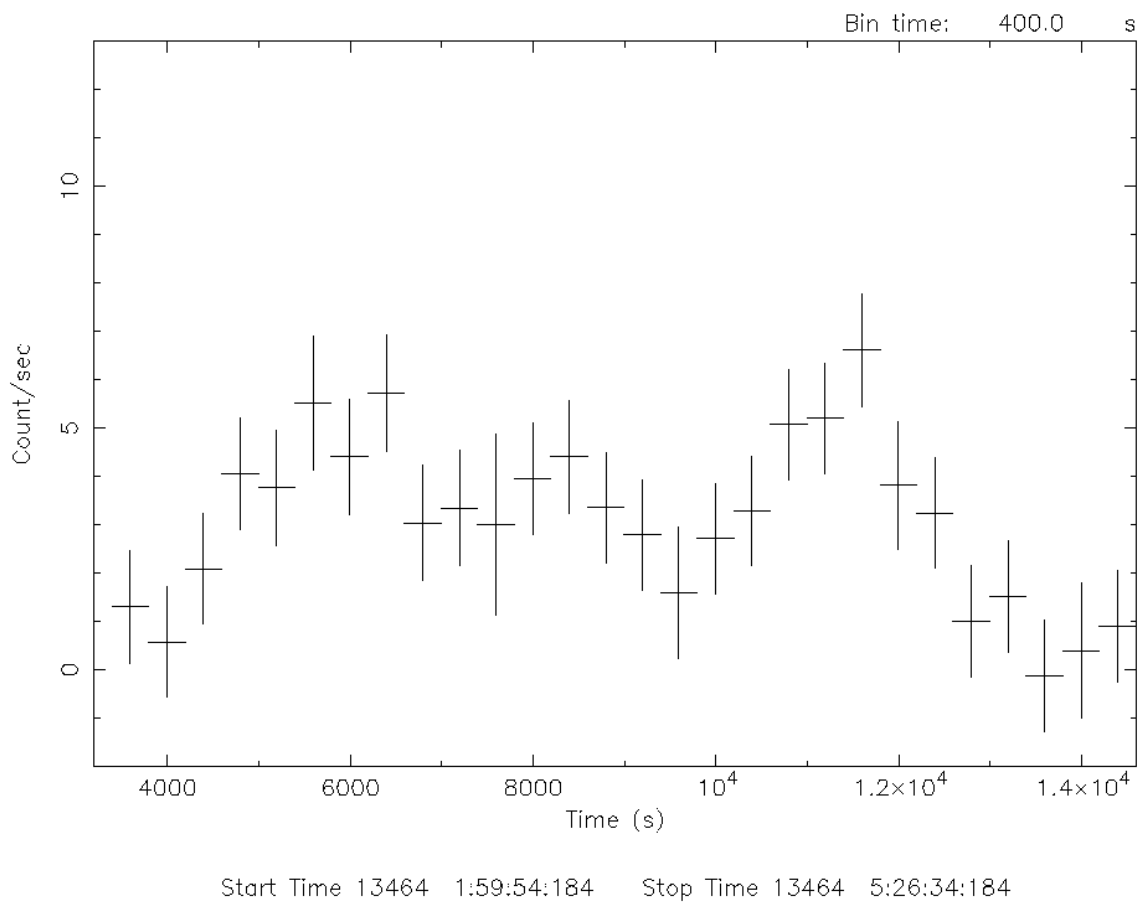}
\caption{The ISGRI light curve (20--60 keV) of a newly discovered outburst of IGR~J16479$-$4514  on 4 April 2005}
\end{figure}

\subsection{XTE~J1743$-$363}
XTE~J1743$-$363 is a faint unidentified X-ray source discovered by RXTE in February 1999 (Markwardt et al. 1999) 
with a flux ranging from 3 to 15 mCrab (2--10 keV); it showed variability on
$\sim$ 1 minute timescales. IBIS detected the source during an observation of the Galactic Center region  performed on 18--20 September 2004,  when 
it was strongly variable with an average flux of 10 mCrab (18--45 keV) (Grebenev \& Sunyaev 2004). A few days later  
a VLA radio observation of XTE~J1743$-$363 did not detect  any strong radio source within the 2$^{'}$ ISGRI error circle either at 4.9 
GHz or at  8.5 GHz (Rupen et al. 2004). Previously, during a $\sim$ 2 million seconds exposure of the Galactic Center region (August--September 2003),
XTE~J1743$-$363 was marginally detected at a flux level of 1.7$\pm$0.2 mCrab (18--60 keV) (Revnitsev et al. 2004).

We report on a newly discovered outburst which unveils for the for time the  fast transient behaviour of XTE~J1743$-$363.
It was detected on 5 October 2004 in the energy range 20--60 keV (Figure 18), with a  duration and peak-flux of $\sim$ 2.5 hours and $\sim$ 40 mCrab (20--60 keV)
respectively.
We extracted a spectrum of the source during the outburst activity (20--60 keV) for which a single
power law gives a reasonably good fit ($\Gamma$=2.9$\pm$0.5, $\chi^{2}_{\nu}$=1.28, d.o.f. 14). A similar fit was also achieved
using  an  optically thin thermal bremsstrahlung model  (kT=22$^{+11}_{-6}$ keV, $\chi^{2}_{\nu}$=1.35, d.o.f. 14).   
\clearpage
\begin{figure}[t!]
\plotone{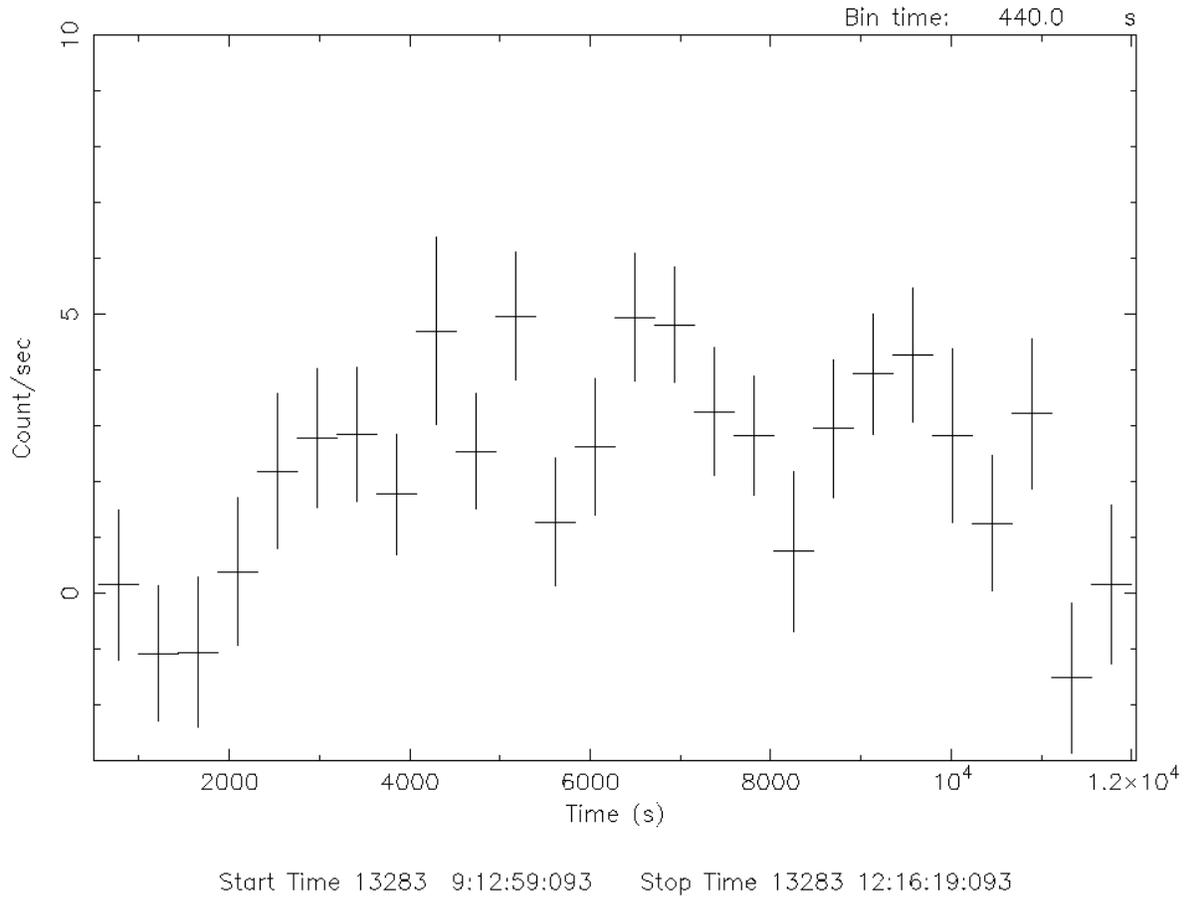}
\caption{The ISGRI light curve (20--60 keV) of a newly discovered outburst of XTE~J1743$-$363.}
\end{figure} 
\clearpage

\subsection{IGR~J16195$-$4945=AX~J161929$-$4945}
IGR~J16195$-$4945 was discovered by INTEGRAL during Core Program observations accumulated between 
27 February and 19 October 2003 (Walter et al. 2004). Subsequently, Sidoli et al. (2005) reported  
the ASCA X-ray source AX~J161929$-$4945 as 
its low energy counterpart. An  ASCA light curve (2--10 keV) of  AX~J161929$-$4945 (Sidoli et al. 2005) 
shows outburst activity which lasted only a few hours, the source was
below the threshold of detectability at the beginning and at the end of the observation.
Moreover, variability on very short timescales is clearly evident with several short  flares.  The ASCA spectrum (1--10 keV)   
is  fitted by an absorbed power law ($\Gamma$ $\sim$ 0.6, N$_H$$\sim$10$^{23}$ cm$^{-2}$ ). 

Analysing public INTEGRAL data, Sidoli et al. (2005) reported on  IBIS  detections of
IGR~J16195$-$4945=AX~J161929$-$4945 in only 2 observations of $\sim$ 1.7 ksec each on 4 and 14 March 2003. In both observations the average flux level was 
$\sim$ 17 mCrab (20--40 keV).

The brightest star in the ASCA error circle (1$^{'}$ radius) is HD 146628 (Sidoli et al. 2005) classified 
as a supergiant of spectral type  B1/B2Ia (distance $\sim$ 7 kpc) in the SIMBAD database. Follow-up observations at X-ray wavelengths
are necessary so as to reduce the error box in order to establish if the supergiant star is really the counterpart of IGR~J16195$-$4945=AX~J161929$-$4945. 

We report a newly discovered fast X-ray outburst that  occurred  
on 26 September 2003.
The duration was $\sim$ 1.5 hours (Figure 19) and it reached a peak-flux of $\sim$ 35 mCrab (20--40 keV).
If we assume HD 146628 to be the optical counterpart of the source, then the 20--40 keV luminosity is $\sim$1.5$\times$10$^{36}$ erg s$^{-1}$.
We extracted a spectrum during the outburst activity. Both black body and  thermal bremsstrahlung models gave 
 unsatisfactory fits ($\chi^{2}_{\nu}$ greater than 1.8).
The best fit model ($\chi^{2}_{\nu}$=1.07, d.o.f. 12) is the sum of  
a black body (kT= 0.8$^{+1.1}_{-0.3}$ keV) with a power law ($\Gamma$=2.78$^{+1.2}_{-1.1}$ ).  
\clearpage
\begin{figure}[t!]
\plotone{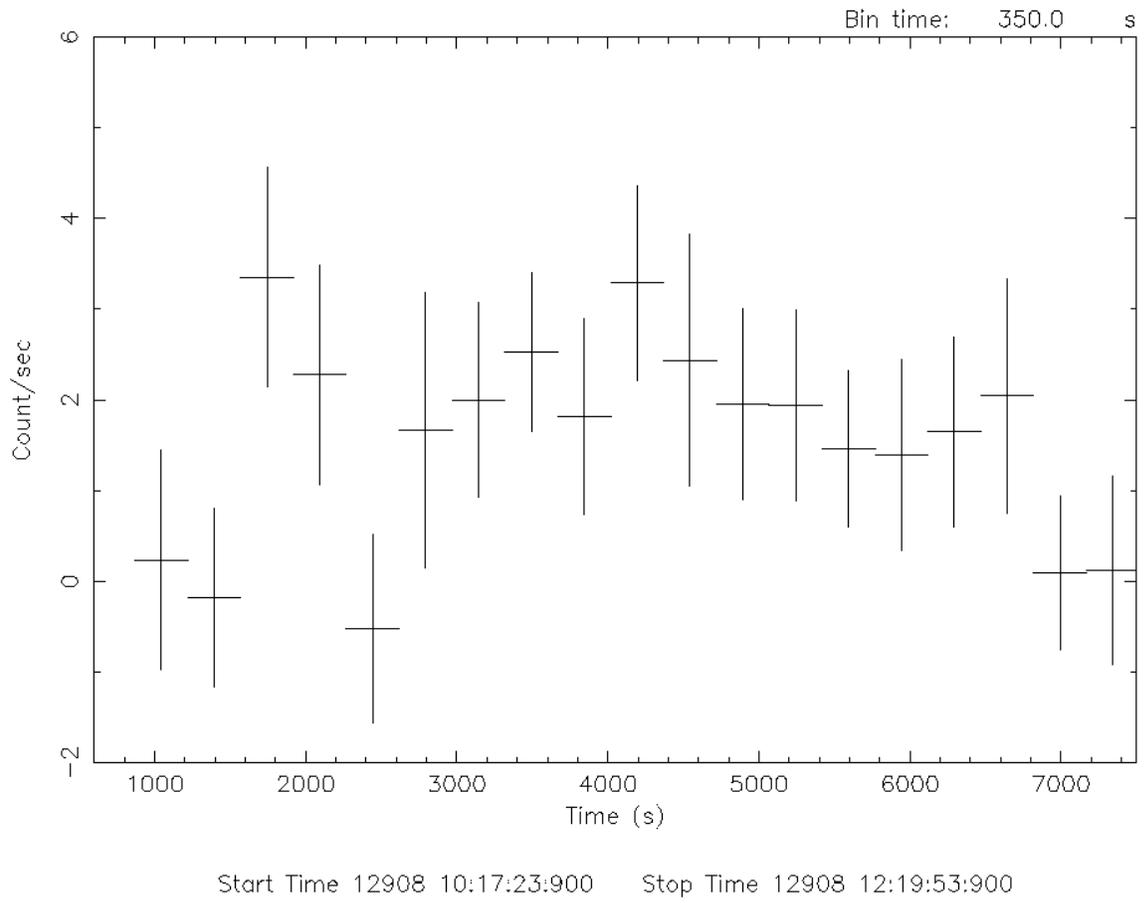}
\caption{The ISGRI light curve (20--40 keV) of a newly discovered outburst of IGR~J16195$-$4945=AX~J161929$-$4945.}
\end{figure} 
\clearpage
\subsection{AX~J1749.1$-$2733}
AX~J1749.1$-$2733 is an unidentified and poorly studied 
X-ray source discovered by ASCA during its survey of the Galactic Center region performed between 1993 and 1999 (Sakano et al. 2002).
The region including this source was observed by ASCA on 6 occasions,  however AX~J1749.1$-$2733 was detected in only 3 of them (September 1996 and 1997, March 1998). 

We report on a newly discovered fast X-ray outburst of AX~J1749.1$-$2733 which unveils for the first time its fast transient nature. 
IBIS detected the source on 9 September 2003 during an outburst lasting $\sim$ 1.3 days (Figure 20), at a position 
(J2000, RA=17$^{h}$ 49$^{m}$ 07.2$^{s}$ DEC=-27$^{\circ}$ 32$^{'}$ 38.4$^{''}$, error circle radius  1.8$^{'}$) which is located 
42$^{''}$ from the ASCA position.
The 20--60 keV peak-flux is $\sim$ 40 mCrab.
A spectrum of the source extracted during this outburst activity  (20--60 keV) can only be reasonably described by the  
($\chi^{2}_{\nu}$=1.46, d.o.f. 12) 
sum of a black body (kT=0.7$^{+0.3}_{-0.1}$ keV) with a power law ($\Gamma$=2.5$\pm$0.2). 
Other spectral models, such as thermal bremsstrahlung, simple power law, black body or Comptonized models, provided  very poor 
fits with  $\chi^{2}_{\nu}$ greater than 3.
\clearpage
\begin{figure}[t!]
\plotone{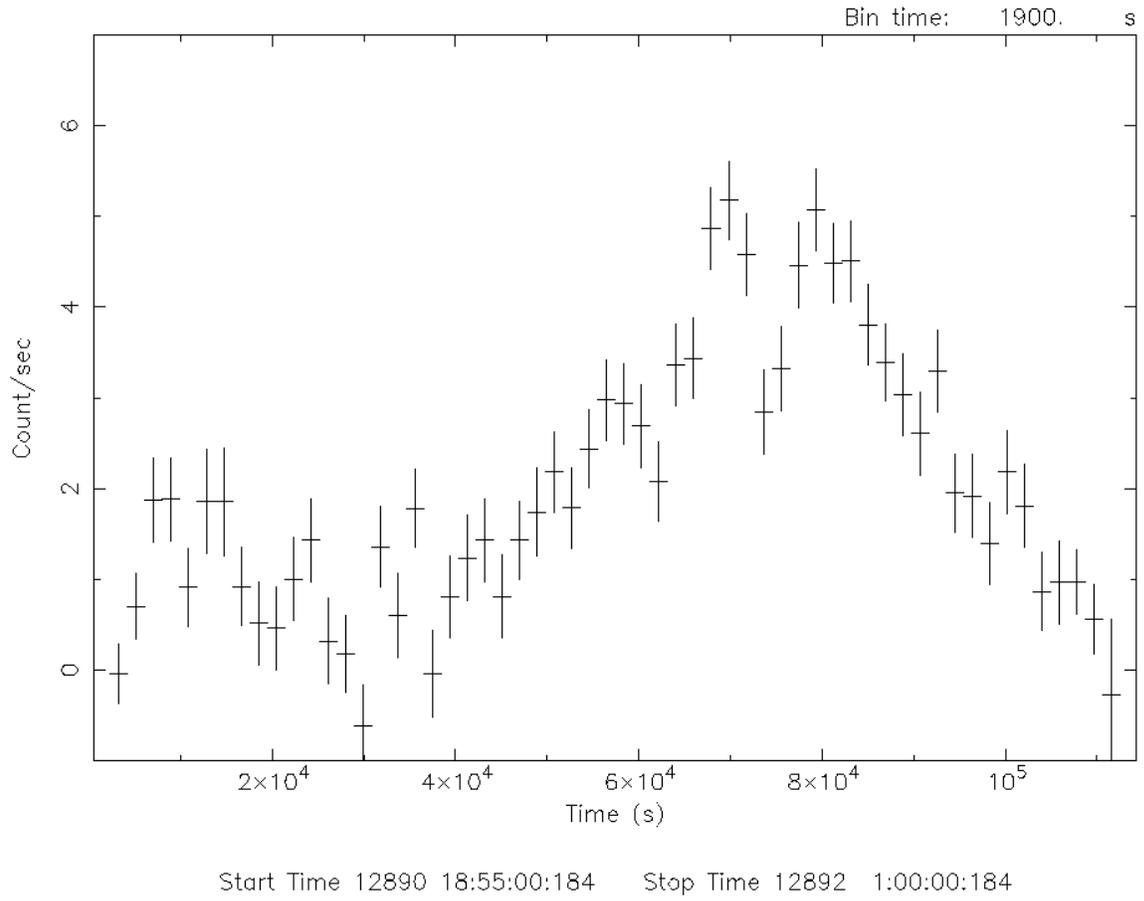}
\caption{The ISGRI light curve (20--60 keV) of a newly discovered outburst of AX~J1749.1$-$2733.}
\end{figure} 
\clearpage

\section{Conclusions}

We have reported on IBIS detections of newly discovered fast X-ray outbursts from 10 sources. Four
of them are SFXTs while the remaining six are candidate SFXTs. Tables 5 and 6 provide a summary of all their known characteristics,  listing 
Galactic coordinates, distance and absorption when available, duration of the outbursts and their peak-flux, peak-luminosity,
temperature of the black body or bremsstrahlung best fit spectrum.    

XTE~J1739$-$302, IGR~J17544$-$2619 and IGR~J18410$-$05355=AX~J1841.0$-$0536 are  known supergiant fast X-ray transients.
The newly discovered fast X-ray outburst from XTE~J1739$-$302 has a peak-flux of $\sim$480 mCrab (20--60 keV), so far it is the strongest  detected 
by IBIS. Newly discovered fast X-ray outbursts from IGR~J18410$-$0535 have been detected in the energy range 20--80 keV and 
to date  IGR~J18410$-$0535 is the only SFXT which had exhibited  a hard tail up to 80 keV.
In the case of IGR~J11215$-$5952, Masetti et al. (2005) recently reported evidence of a  supergiant HMXB nature. We strengthen this association 
since we unveiled for the first time its fast X-ray transient nature. 

As for the 6 unidentified X-ray sources reported in this paper,  we have presented newly discovered fast outbursts 
which strongly resemble those of known SFXTs.
Although the optical counterparts have not yet been identified,
they are candidate SFXTs.
In particular,  IGR~J16479$-$4514 is  known to display fast X-ray outbursts 
which have been detected by IBIS (Sguera et al. 2005). Concerning
the remaining 5 unidentified X-ray sources (AX~J161929$-$4945, IGR~J16418$-$4532, XTE~J1743$-$363,
AX~J1749.1$-$2733, IGR J17407$-$2808), 
we reported  their fast X-ray transient nature for the first time. Among them, IGR~J17407$-$2808 showed a peculiar fast 
transient activity lasting only a few minutes and we cannot rule out the possibility that its fast outburst is not from a supergiant HMXB.
The timing behaviour resembles the so called burst-only sources detected by the WFCs on board BeppoSAX (Cornelisse et al. 2004). 
It is worth noting  that 2 more sources have been reported in the literature as candidate SFXTs,
SAX~J1818.6-1703 (Sguera et al. 2005, Grebenev \& Sunyaev 2005, Negueruela et al. 2005a) 
and XTE~J1901+014 (Negueruela et al. 2005a).

In Figure 21 we show  the angular distribution off the Galactic Plane of the 10 sources reported in this paper (filled black bars) and of 
all HMXBs (empty white bars) reported in the 2nd IBIS/ISGRI gamma-ray catalog (Bird et al. 2006). As far as we can tell from the limited 
statistics, the density of the 10 sources reported in our paper appears to be greater towards the Galactic plane, 
with 9 of the 10 having $|$b$|$$<$ 1.5$^{\circ}$.  This is to be expected if they are supergiant
HMXBs, due to the very young ages of their progenitor stars. 
If we assume that INTEGRAL is observing the Galactic Plane randomly and 
combine the information provided in section 2 (see Figure 2) with the number of  SFXTs detected by IBIS/ISGRI,
then we estimate a rough rate of detection of SFXTs equal to $\sim$ 20 per year.

Although SFXTs  are difficult to detect, the number known  is rapidly increasing thanks to INTEGRAL observations.
Apart from the SFXTs mentioned in this paper,  two more have been reported in the literature;
IGR~J16465$-$4945 (Negueruela et al. 2005a) and AX~J1845.0-0433 (Yamauchi et al. 1995, Negueruela et al. 2005a). 
On the contrary there are 8 candidate SFXTs and if just a few of them could be confirmed, then their number 
would be already comparable to that of classical supergiant HMXBs. Moreover, the class of SFXTs  could be much larger  than the previous  14 cited objects. 
A population of still undetected supergiant fast X-ray transient sources could be hidden in our Galaxy. 
Ongoing observations with INTEGRAL may yield further detections of such sources.

\begin{figure}[t!]
\plotone{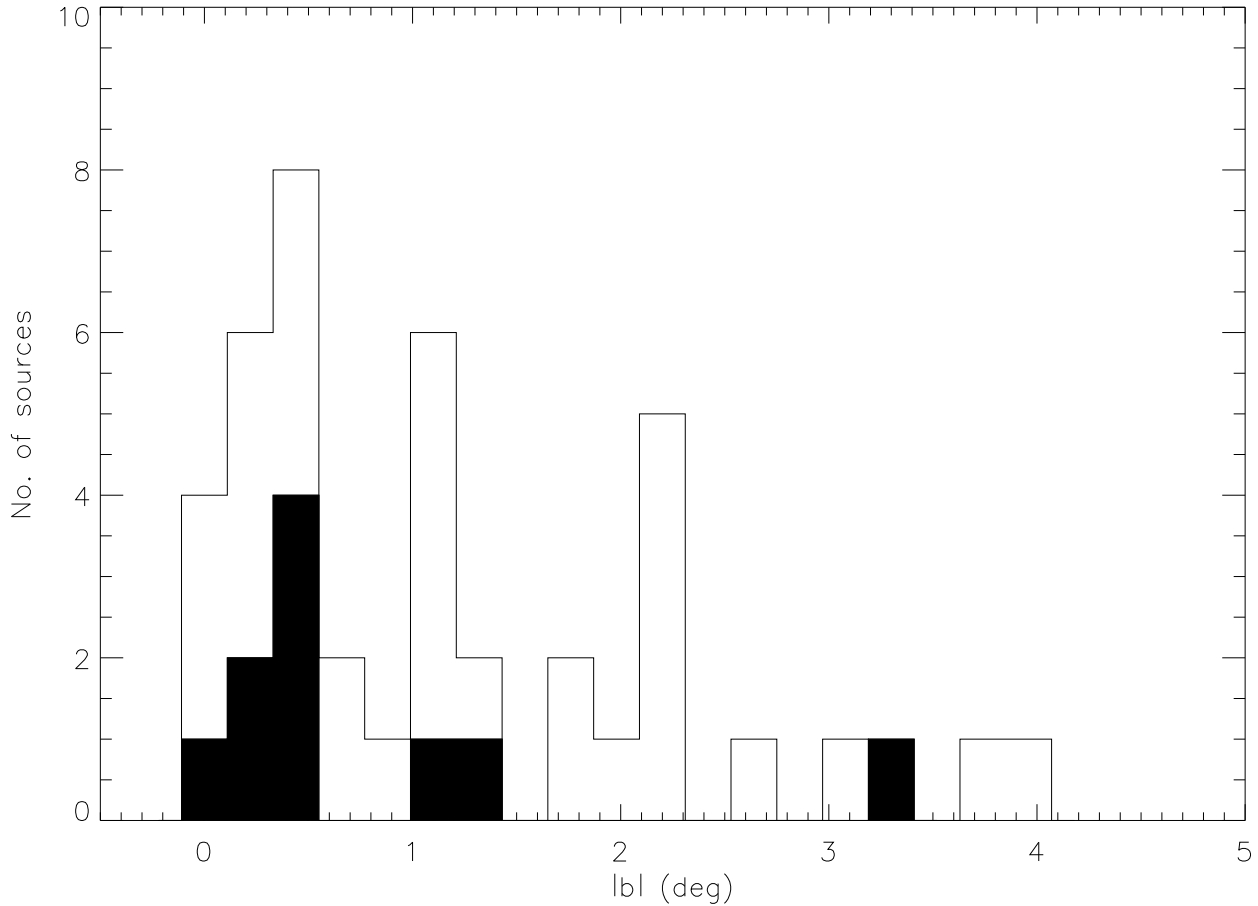}
\caption{The angular distribution off the Galactic Plane of the 10 sources reported in this paper (filled black bars) and of  
all HMXBs (empty white bars) as taken from the  2nd IBIS/ISGRI gamma-ray catalog (Bird et al. 2006).}
\end{figure}

\clearpage
\rotate
\begin{table}[t!]
\tiny
\caption {Summary of characteristics of the 4 supergiant fast X-ray transients}
\begin{tabular}{ccccccccccc}
\hline
\hline
Source  & l & b & distance &  N$_{H}$  & Date  & duration & flux at the peak & Luminosity-peak & kT$_{BB}$ &  kT$_{BR}$   \\
 & & & kpc &  $\times$10$^{22}$ cm$^{-2}$  & & h & mCrab &  $\times$10$^{36}$ erg s$^{-1}$     & keV & keV\\
\hline
\hline
XTE~J1739$-$302  & 358.07 & 0.45 & 2.3   & 3--38 & 22/03/2003 & $\sim$ 2  & $\sim$ 250  (20--30 keV)  & 0.7 (20--30 keV)& & \\
                                                    &    & &    &  &  26/08/2003  & $\sim$ 14 & $\sim$ 120  (18--60 keV) & & &   \\  
                     &        &      &              & & 6/09/2003 &  $\sim$ 7 &  $\sim$ 60  (18--60 keV)& & & \\
                     & & & &               & 9/03/2004 &  $\sim$ 0.5 &  $\sim$ 150  (20--30 keV)  &  & & \\
                                     & & &              &                 & 10/03/2004 &  $\sim$ 1.5 &  $\sim$ 250  (20--30 keV)  & & & \\
                       & &        &                 &              & 21/08/2004  &  $\sim$ 3 &  $\sim$ 480  (20--60 keV)  &  3.5 (20--60 keV) & &  \\
IGR~J17544$-$2619 & 3.24 & -0.34 & 3 &   $\sim$ 3  & 17/09/2003 01h  & $\sim$ 2  & $\sim$ 400  (20--40 keV)  & 3.2 (20--40 keV) & & \\
 & & & &               &  17/09/2003 06h & $\sim$ 8 & $\sim$ 80 (25--50 keV) & &   & \\
 & & & &               &   8/03/2004 &  $\sim$ 10 &  $\sim$ 240  (20--60 keV)&  3 (20--60 keV) &  4.4$\pm$0.25  & 9.5$\pm$0.9 \\
 & & & &               & 21/09/2004 &  &  $\sim$ 70  (20--40 keV)  & 0.57 (20--40 keV) & &   \\
 & & & &               & 12/03/2005 &  $\sim$ 0.5 &  $\sim$ 150  (20--30 keV)  &  & 2.9$\pm$0.4  & 5$\pm$1\\ 
IGR~J18410$-$0535  & 26.76 & -0.239 & & $\sim$ 6 &  24/04/2003 & $\sim$ 7 & $\sim$ 120  (20--80 keV) &  & 8$\pm$0.9 & \\                  
 & & & &               & 6/05/2003 & $\sim$ 1.1 & $\sim$ 120  (20--80 keV) &  & 9$\pm$0.9 &   \\
  & & & &              & 8/10/2004 &   &   &   & &    \\
IGR~J11215$-$5952  &  291.89 & 1.07 & 6.2 & & 4/07/2003 & $\sim$ 3 & $\sim$ 40 (20--40 keV) & 1.4 (20--40 keV) & 6.2$\pm$0.6 & 19$^{+5}_{-3.5}$ \\
 & & & &               &  22/04/2005 & & & & & \\
\hline
\hline
\end{tabular}

\begin{center}
\caption {Summary of characteristics of the 6 candidate supergiant fast X-ray transients}
\begin{tabular}{ccccccccccc}
\hline
\hline
Source  & l & b & distance &  N$_{H}$  & Date  & duration & flux at the peak & Luminosity-peak & kT$_{BB}$ &  kT$_{BR}$   \\
 & & & kpc &  $\times$10$^{22}$ cm$^{-2}$  & & h & mCrab &  $\times$10$^{36}$ erg s$^{-1}$     & keV & keV\\
\hline
\hline
IGR~J17407$-$2808 & 0.12 & 1.34 & & & 9/10/2004 & $\sim$ 3 minutes &  805 (20--60 keV) &&  7$\pm$0.7  & 23$^{+7}_{-4.5}$ \\
IGR~J16479$-$4514 & 340.14 & -0.12 &  & $\sim$ 12 &  5/03/2003 & $\sim$ 3.5  & $\sim$ 850 (20--30 keV)    & & & \\
 & & & &               &   28/03/2003 & $\sim$ 1.5 & $\sim$ 40$\dagger$ & & & \\  
& & & &               & 21/04/2003 &  $\sim$ 0.5 &  $\sim$ 160$\dagger$ & & & \\
& & & &               &  8--10/08/2003 &  &  &  & & \\
& & & &               &   14/08/2003 & $\sim$ 2 &  $\sim$ 44$\dagger$ & & & \\
& & & &               &  7/09/2004 &  $\sim$2 &  $\sim$ 125  (20--60 keV)  & &   &   \\
& & & &               &  16/09/2004 &  $\sim$ 2.5 &  $\sim$ 120 (20--60 keV)  &  & 7.4$\pm$0.5 & \\
& & & &               &  4/04/2005 &  $\sim$ 2.5 &  $\sim$ 60  (20--60 keV)  & & & \\
IGR~J16418$-$4532     & 339.18 & 0.5 & & & 1--5/02/2003 &   &  & & & \\
& & & &               &  26/02/2004   & $\sim$ 1 & $\sim$ 80 (20--30 keV) & & &  \\
IGR~J16195$-$4945     & 333.54 & 0.33 &  & $\sim$ 12 & 4/03/2003 & $\sim$ 0.5  &  $\sim$ 17$\star$& & \\
& & & &               &  14/03/2003 & $\sim$ 0.5  & $\sim$ 17$\star$   & & \\
& & & &               &  26/09/2003 & $\sim$ 1.5 & $\sim$ 35 (20--40 keV) & & & \\
XTE~J1743$-$363         & 353.39 & -3.4 & & & 5/10/2004 & $\sim$ 2.5 & $\sim$ 40 (20--60 keV) & &  & 22 $^{+11}_{-6}$  \\   
AX~J1749.1$-$2733     & 1.585 & 0.051 & & & 9/09/2003 & $\sim$ 31 & $\sim$ 40 (20--60 keV) & & & \\
\hline
\hline
\end{tabular}
\end{center}
$\dagger$ = Average flux (20--30 keV) during the outburst\\
 $\star$  = Average flux (20--40 keV) during the outburst\\
\end{table}  

\clearpage

\acknowledgements
We thank the anonymous referee for very useful comments which helped us to improve the paper.
This research made use of data obtained from the HEASARC and  SIMBAD database. This research 
has been supported by  University of Southampton School of Physics and Astronomy. AB, PU, AM, LB, JBS acknowledge the ASI financial
support via grant I/R/046/04.

\end{document}